\documentclass[fleqn,usenatbib,useAMS]{mnras}
\newcommand{\bw}{}
\usepackage[T1]{fontenc}
\usepackage{ae,aecompl}

\usepackage{graphicx}	% Including figure files
\usepackage{amsmath}	% Advanced maths commands
\usepackage{amssymb}	% Extra maths symbols
\usepackage{pdflscape}
\usepackage{dsfont}

\usepackage[usenames,dvipsnames]{xcolor}          % For color text: \color command
\usepackage{framed}

\DeclareFontFamily{OT1}{pzc}{}
\DeclareFontShape{OT1}{pzc}{m}{it}%
             {<-> s * [1.1500] pzcmi7t}{}
\DeclareMathAlphabet{\mathscr}{OT1}{pzc}%
                                 {m}{it}

\hypersetup{pdfauthor={P. S. Cally},
               pdftitle={Alfv\'en Waves in the Structured Solar Corona},
               pdfkeywords={Sun: oscillations --  magnetohydrodynamics -- Sun: helioseismology -- Sun: corona},
               bookmarksnumbered=true}
   \volume{{\rm in press}}

\newcommand{\half}{\frac{1}{2}}

\newcommand{\re}{\mathop{\rm Re}\nolimits}
\newcommand{\im}{\mathop{\rm Im}\nolimits}

\newcommand{\rme}{\mathrm{e}}
\newcommand{\rmi}{\mathrm{i}}

\newcommand{\e}{\hat{\bmath{e}}}

\newcommand{\bk}{\bmath{k}}

\newcommand{\boldv}{{\bmath{v}}}

\newcommand{\x}{\bmath{x}}

\newcommand{\B}{\mathbf{B}}

\newcommand{\F}{\mathbf{F}}

\renewcommand{\P}{\mathbf{P}}

\newcommand{\vdot}{{\boldsymbol{\cdot}}}
\newcommand{\vcross}{{\boldsymbol{\times}}}
\newcommand{\grad}{\mbox{\boldmath$\nabla$}}

\newcommand{\diag}{\mathop{\rm diag}}
\newcommand{\thth}{\hspace{1.5pt}}

\newcommand{\Curl}{\grad\vcross\thth}

\newcommand\Div{\grad\vdot\thth}

\newcommand{\kperp}{k_{\scriptscriptstyle\!\perp}}

\newcommand{\kpar}{k_{\scriptscriptstyle\parallel}}

\newcommand{\ri}{\mathrm{i}}

\renewcommand{\leq}{\leqslant}  \renewcommand{\le}{\leqslant}
\renewcommand{\geq}{\geqslant}  \renewcommand{\ge}{\geqslant}

\title[Alfv\'en Waves in the Structured Solar Corona]{Alfv\'en Waves in the Structured Solar Corona}

\author[P.S.~Cally]{
Paul S.~Cally\thanks{E-mail: paul.cally@monash.edu}
\\
School of Mathematical Sciences and Monash Centre for Astrophysics, Monash University, Clayton, Victoria 3800, Australia}

\date{}
\pubyear{2015}

\begin{document}
\label{firstpage}
\pagerange{\pageref{firstpage}--\pageref{lastpage}}
\maketitle

\begin{abstract}
A simple model of a periodic ensemble of closely packed flux tubes sitting atop a vertically stratified layer reveals that an incident fast wave from below preferentially converts almost immediately to Alfv\'en waves in the flux tubes, with kink waves restricted to at most a very few Fourier modes.  This suggests that observations of coronal kink modes in such structured systems may greatly underestimate the net wave energy flux being transported into and through the corona, much of which may reside in harder-to-observe Alfv\'en waves. The processes of mode conversion/resonant absorption and Alfv\'en phase mixing are implicated. It is suggested that the Sun's internal $p$-mode field -- the 5-minute oscillations -- may contribute substantially to the process by supplying incident fast waves in the chromosphere that scatter and mode-convert in the tube ensemble.
\end{abstract}


%%%%%%%%%%%%%%%%%%%%%%%%%%%%%%%%%%%%%%%%%%%%%%%%%%%%%

\section{Introduction}%
Since Alfv\'en waves were first shown to be incompressive solutions of the magnetohydrodynamic (MHD) equations in a conducting fluid \citep{Alf42aa}, they have been postulated to contribute to many phenomena on the Sun. In modern times, their role from photosphere to solar wind has been widely invoked \citep{Cravan05aa}, including as a source of coronal heating via a turbulent cascade and of solar wind acceleration \citep{CravanEdg07aa}.

The first direct detection of coronal Alfv\'en waves was reported a decade ago by \citet{TomMcIKei07aa} in an article pointedly entitled \emph{Alfv\'en Waves in the Solar Corona}: \emph{``We report the detection of Alfv\'en waves in intensity, line-of-sight velocity, and linear polarization images of the solar corona taken using the FeXIII 1074.7-nanometer coronal emission line with the Coronal Multi-Channel Polarimeter (CoMP) instrument at the National Solar Observatory''}. At only 0.3 $\rm km\,s^{-1}$ rms amplitude though, these Alfv\'en waves are too weak to be energetically important. The observed velocity fluctuations show a distinct signature of the solar $p$-mode 5-minute oscillations, suggesting a link with the Sun's internal seismology. This is absent in intensity though, indicating that the helioseismology-related power is predominantly in incompressive or near-incompressive modes, and specifically not in slow modes.

This was quickly followed by Hinode Solar Optical Telescope (SOT) CaII H-line observations of swaying chromospheric spicules by \citet{De-McICar07aa}, who judged these Alfv\'en waves \emph{``strong enough to power the solar wind''}. Their sophisticated 3D radiative MHD simulations confirmed the interpretation. They \emph{``(did) not see evidence for stable waveguides or MHD kink-mode waves''} in the simulations. Importantly, transmission coefficients across the chromosphere-corona transition region (TR) of 3--15\% were calculated (compared to 5\% reported by \citealt{Cravan05aa}), suggesting that the corona and solar wind may indeed benefit from a substantial Alfv\'en energy injection. 

\citet{CirGolLun07aa} found similar Alfv\'en waves in longer-lived solar X-ray jets, suggesting a ubiquitous phenomenon.

However, \citet{ErdFed07aa} in the same Special Section of \emph{Science} raised the question of whether these ``Alfv\'en waves'' were instead kink waves. This was on the basis that the only allowable polarization of Alfv\'en waves in a radially structured cylindrical flux tube is torsional, and not transverse as observed. Torsional Alfv\'en waves on such tubes would show up only as line broadening. \citet{VanNakVer08aa} also argued that these waves are ``guided kink magnetoacoustic waves'', not Alfv\'en waves. 

\citet{JesMatErd09aa} note the kink/Alfv\'en uncertainty, but concentrate specifically on pure torsional oscillations in axisymmetric flux tubes that are undoubtedly Alfv\'en \citep[see also][]{KudShi99aa}. \citet{MatJesErd13aa} argued that, although fundamentally magnetoacoustic, transversal kink waves are only very weakly compressional, and so share many characteristics with true Alfv\'en waves.

Coincidently, the discrete flare-induced coronal loop oscillations observed with the Transition Region and Coronal Explorer (TRACE) were interpreted as kink waves from the first \citep{AscFleSch99aa,Ascde-Sch02aa,NakOfmDel99aa},
though with their observed decay soon being ascribed to resonant absorption, i.e., conversion to Alfv\'en waves at resonant surfaces \citep{GooAndAsc02aa}. These and other coronal wave types (e.g., EIT waves, compressible waves in plumes and loops, and various types of standing loop waves) were interpreted as MHD waves of one type or another (a detailed discussion distinguishing the various MHD wave types in magnetic flux tubes is presented by \citealt{GooTerAnd09aa}). These observations and their interpretations are discussed in historical and scientific detail in the \emph{Living Review} by \citet{NakVer05aa}, and need not be revisited here.

Though fully aware of the plethora of observations of coronal waves identified as magneto-acoustic (the fast and slow waves, including their kink and sausage manifestations in flux tubes), \citet{TomMcI09aa}
made the case that the ``spatially and temporally ubiquitous'' waves seen in the CoMP Doppler time series by \citet{TomMcIKei07aa} are novel in character. This is because (i) they lack appreciable intensity fluctuations (suggesting Alfv\'en or near-Alfv\'en character), making them invisible to intensity imaging instruments; (ii) their displacement amplitudes are nearly an order of magnitude below what could be observed by TRACE, and two orders of magnitude below SOHO/EIT capabilities; and (iii) existing Doppler imaging instruments lacked the required sensitivity, spatial extent, and cadence. They claimed the CoMP observations could widen the remit of coronal seismology. 

These ubiquitous waves in structured ensembles of flux tubes will be modelled in subsequent sections, with particular attention to irrotational and incompressive parts, which display respectively kink-like and Alfv\'en-like characteristics.

\citet{McIde-Car11aa} report far higher coronal Alfv\'en wave amplitudes than \citet{TomMcIKei07aa}, 20 $\rm km\,s^{-1}$ rather than 0.3 $\rm km\,s^{-1}$, sufficient to power the quiet solar corona and fast solar wind. Higher spatial and temporal resolution available using the He II 304-{\AA} and Fe IX 171-{\AA} channels of the Atmospheric Imaging Assembly (AIA) aboard the Solar Dynamics Observatory (SDO) allowed direct imaging of swaying motions rather than predominantly line broadening seen with CoMP, reducing the effect of high-optical-depth averaging. By this stage, the waves were being referred to as ``Alfv\'enic'' rather than ``Alfv\'en'', allowing for them being hybrid Alfv\'en/kink in nature. 

There are differences between Alfv\'en and kink waves that may have practical consequences. Although only weakly compressional, kink waves are to some extent subject to dissipation mechanisms associated with compressibility, and Alfv\'en waves are not. The primary source of kink wave dissipation in structured flux tubes though is believed to be resonant absorption, which is simply mode conversion to Alfv\'en waves \citep{CalAnd10aa} tightly bound to resonant surfaces (for a single frequency). In the radially stratified flux tube context, these Alfv\'en waves are predominantly torsional  and have far reduced observational signatures, so by this stage the kink wave may appear to have died out \citep{GooAndAsc02aa}. The Alfv\'en waves generated in this way are subject to phase mixing (cascade to smaller scales) and therefore ultimately to dissipation via non-ideal processes.

\citet{GooSolTer14aa} make the point that the identification of kink and Alfv\'en waves in structured plasmas is not absolute or global. With particular reference to nonuniform circular-cross-section flux tubes, they argue that the oscillations are hybrid in nature, taking on either classic kink-like or torsional-Alfv\'en characteristics in different spatial regions or time periods (in an initial value problem). In particular, the oscillations become more Alfv\'en-like (Alfv\'enic) near the Alfv\'en resonance layers (in agreement with the modelling of \citealt{CalAnd10aa} and \citealt{HanCal11aa}). This view is supported by the results presented herein.

Photosphere-to-heliosphere Alfv\'en models commonly invoke the photospheric convective power spectrum as the source of Alfv\'en waves. However, with the very low ionization fraction of the temperature minimum region taken into account, \citet{VraPoePan08aa} conclude that the generated Alfv\'en flux is reduced by orders of magnitude, which is potentially fatal for such atmospheric Alfv\'en wave models.

An alternate source of Alfv\'en waves that does not suffer this weakness relies on fast-to-Alfv\'en mode conversion in the chromosphere, where the ionization fraction is much higher and standard MHD more applicable. With a given horizontal wavenumber, an upward travelling fast wave in a vertically stratified atmosphere reflects where its horizontal phase speed matches the Alfv\'en speed (assuming a zero-$\beta$ (cold) plasma), and partially converts to upward or downward travelling Alfv\'en waves \citep{CalHan11aa}. The same behaviour is seen in the warm plasma model of \citet{CalGoo08aa}, and verified in simulations by \citet{KhoCal11aa,KhoCal12aa}, and \citet{Fel12aa}. 

In this scenario internal solar $p$-modes (partially) convert to fast waves at the Alfv\'en acoustic equipartition surface \citep{SchCal06aa}, and thence to Alfv\'en waves higher up in the low-$\beta$ upper chromosphere. \citet{HanCal12aa} found that these newly created Alfv\'en waves were better able to penetrate the transition region than those originating from the photosphere, with transmission coefficients up to 30\%, depending on field inclination and wave attack angle. This fast-to-Alfv\'en conversion process is essentially the same as the resonant absorption mechanism in flux tubes mentioned above, but with the required Alfv\'en speed gradient being a consequence of gravitational stratification instead of radial tube structure. For inclined magnetic field though, the generated Alfv\'en waves are spatially distributed rather than confined to discrete resonant surfaces \citep{CalHan11aa}. Linking atmospheric Alfv\'en waves to internal seismology is in line with the identification of peak power with the 5-minute $p$-mode spectrum by \citet{TomMcIKei07aa}.

These mode conversion analyses though assume either vertical or predominantly vertical Alfv\'en speed stratification with little or no cross-field structure. The recent observations on the other hand have been in the context of a very cross-field-structured corona. In the photosphere and low chromosphere, the density scale height is typically 100--150 km, which is much shorter than most relevant horizontal length scales, so vertically stratified models are appropriate. On the other hand, moving higher in the atmosphere sees the magnetic field take over from gravity as the primary cause of inhomogeneity, either directly or by allowing different flux tubes to contain different temperature and density plasma. Both scenarios present opportunities for mode coupling. It is the purpose of this article to explore a model in which the atmosphere transitions smoothly from vertical to cross-field inhomogeneities with increasing height $z$, and to calculate the relative wave energy fluxes carried by fast and Alfv\'en waves as a function of $z$.

Although \citet{GooTerAnd09aa} specifically reject calling kink waves ``fast'', \citet{De-Nak12aa} say \emph{``the kink mode is locally a fast magnetoacoustic wave, propagating obliquely to the magnetic field and guided along the field-aligned plasma structure (a waveguide) by reflection or refraction''}. That this is indeed the case is made clear by \citet{PasWriDe-10aa,PasWriDe-11aa}; the reader is particularly referred to the cartoon Figure 1 of the latter article. This view of the kink (or sausage, or fluting) mode as a fast wave trapped in a low Alfv\'en speed waveguide will inform the interpretation of the numerical solutions obtained in the following sections.

So, the precise definition of ``fast'' and ``Alfv\'en'' in an inhomogeneous plasma is not clear-cut. In general it is not even possible to separate them unambiguously, as the two mode types are inextricably coupled (though if the coupling is weak, they may be separated as zeroth order modes in a perturbation expansion). Nevertheless, some quantification can be given in terms of irrotational and incompressive parts of the displacement vector, that to some extent may be associated primarily with the fast and Alfv\'en parts. To separate them unambiguously, a uniform region will be appended at the top of the computational region so that the modes decouple. 

It will be shown that the easily-observed kink-like irrotational part typically carries far less wave-energy flux upward than does the difficult-to-observe Alfv\'en-like incompressible part. This suggests that observations may significantly understate the true MHD wave flux into the corona.

%%%%%%%%%%%
\section{Model and Mathematical Development}
For the purposes of this study, it will be sufficient to adopt the zero-$\beta$ approximation (cold plasma model), in which the sound speed is assumed negligible compared to the Alfv\'en speed. This is a reasonable description of the regime of interest in the solar corona.

\subsection{Fundamental Equations} \label{fund}
Consider a cold ideal MHD plasma with uniform magnetic field $\B_0$, supporting both fast and Alfv\'en waves. The slow wave has been frozen out by the cold plasma assumption $c/a\to0$, where $c$ and $a$ are the sound and Alfv\'en speeds respectively. As shown by \citet{CalHan11aa}, the plasma displacement $\bxi$ obeys the linearized wave equation
\begin{equation}
\left(\upartial_\parallel^2-\frac{1}{a^2}\upartial_t^2\right)\bxi  =-\grad_{\!\text p}\chi
,                               \label{basiceqn}
\end{equation}
where $\chi=\Div\bxi$ is the dilatation, $\upartial_t$ is the time derivative, $\upartial_\parallel=\e_\parallel\vdot\grad$ is the field-aligned directional derivative, $\e_\parallel=\hat\B_0$ is the unit vector in the direction of the magnetic field, and $\grad_{\!\text p}=\grad-\e_\parallel\upartial_\parallel$ is the complementary perpendicular component of the gradient. Note that there is no displacement along field lines $\B_0\vdot\thth\bxi=0$. Even though the magnetic field $\B_0$ is assumed uniform, the square of the Alfv\'en speed $a^2=B_0^2/\umu_0\rho_0$ is an arbitrary function of position through its dependence on the density $\rho_0(\x)$.

The displacement may be Helmholtz-decomposed into irrotational and incompressive parts using potentials,
\begin{equation}
\begin{split}
 \bxi &=\grad_\text{p}\Phi-\Curl\Psi\e_\parallel=\bxi_\text{f}+\bxi_\text{A}   \\[4pt]
 &= \left(\upartial_\perp\Phi-\upartial_y\Psi\right)\e_\perp+\left(\upartial_y\Phi+\upartial_\perp\Psi\right)\e_y , 
\end{split} \label{ptl}
\end{equation}
whence
\begin{equation}
\chi=\nabla_\text{p}^2\Phi \quad\text{and}\quad\zeta=\nabla_\text{p}^2\Psi,
\label{Poissons}
\end{equation}
where $\zeta=\e_\parallel\vdot\Curl\bxi=\Div(\bxi\thth\vcross\thth\e_\parallel)$. The potential
$\Phi$ represents the fast wave, and $\Psi$ characterizes the Alfv\'en wave in a uniform medium. 

Equation (\ref{basiceqn}) then becomes
\begin{subequations}\label{PhiPsi}
\begin{align}
\left(\nabla^2-\frac{1}{a^2}\upartial_t^2\right)\upartial_\perp\Phi &= 
\left(\upartial_\parallel^2-\frac{1}{a^2}\upartial_t^2\right)\upartial_y\Psi   \\[4pt]
\left(\upartial_\parallel^2-\frac{1}{a^2}\upartial_t^2\right)\upartial_\perp\Psi &=
-\left(\nabla^2-\frac{1}{a^2}\upartial_t^2\right)\upartial_y\Phi, 
\end{align}
\end{subequations}
neatly separating the two components in terms of the fast and Alfv\'en operators $\mathcal{F}=\nabla^2-a^{-2}\upartial_t^2$ and $\mathcal{A}=\upartial_\parallel^2-a^{-2}\upartial_t^2$, showing how they couple. As previously noted by \citet{CalHan11aa}, the fast and Alfv\'en waves decouple in the two-dimensional (2D) case $\upartial_y\equiv0$. They also decouple if the Alfv\'en speed $a$ is uniform, for which case $\mathcal{F}\chi=0$ and $\mathcal{A}\zeta=0$ result.

Boundary conditions are chosen to model mode conversion from fast waves injected at the bottom, $z_{\text{bot}}$, with no incoming Alfv\'en waves there, and with only evanescent or escaping waves at the top, $z_{\text{top}}$. Details are presented in Section \ref{BCs}.

Of course, there is a gauge ambiguity about the potentials $\Phi$ and $\Psi$: the physical displacement $\bxi$ is invariant under the mapping $\Phi\to\Phi+\upartial_y\Upsilon$, $\Psi\to\Psi+\upartial_\perp \Upsilon$ for arbitrary perpendicular-harmonic function $\Upsilon$ (i.e., $\nabla_\text{p}^2\Upsilon=0$). This produces an uncertainty in partitioning the displacement into fast and Alfv\'en parts, but none in $\chi$ or $\zeta$.

%%%%%%%%%%%%%%%%%%%%%%%%%%%%%%%%%%%%%%%%%%%%%%%%%
\subsection{What are Fast and Alfv\'en Waves Anyway?}
The common understanding of fast and Alfv\'en waves derives from the well-known case of a uniform atmosphere, where they may be unambiguously distinguished. Specifically, in that circumstance,
\begin{enumerate}
\item They each have their own dispersion relations, $\omega^2=a^2 k^2$ and $\omega^2=a^2 k_\parallel^2$ respectively (in a zero-$\beta$ plasma);
\item The group velocity (energy propagation vector) of the fast wave is aligned with the wave vector whilst that of the Alfv\'en wave is identically along the magnetic field direction; 
\item The fast wave is irrotational and the Alfv\'en wave is incompressive; and 
\item Their displacements are orthogonal to each other. 
\end{enumerate}
Which of these characteristics is essential, and which is incidental and specific to the uniform plasma case? This goes to the definition of the fast and Alfv\'en waves in a structured medium.

The distinction adopted here via Equation (\ref{ptl}) is based on the fast wave being irrotational and the Alfv\'en wave incompressive. This makes $\chi$ and $\zeta$ unambiguously characteristic of the fast and Alfv\'en wave respectively. When there is an ignorable direction, $\upartial_y\equiv0$, Equations (\ref{PhiPsi}) confirm that this definition also recovers the expected ``dispersion relations'' that $\Phi$ is associated exclusively with the fast wave operator $\mathcal{F}$ and $\Psi$ with the Alfv\'en operator $\mathcal{A}$. In this case, the polarizations (in the $\perp$ and $y$ directions respectively) are orthogonal too.

Ultimately, in the general case without an ignorable direction, it is orthogonality that is jettisoned. This is crucial, as it allows the two modes to interact and exchange energy. In that sense, they are no longer distinct, and alternate partitions are feasible. Strictly, there are no \emph{pure} fast and Alfv\'en waves anymore, though as a matter of definition, the terms ``fast'' and ``Alfv\'en'' will continue to be applied to the $\Phi$ and $\Psi$ solutions respectively. Especially where the coupling is weak, this is a useful convention. 

However, the convention adopted is irrelevant in a uniform region appended at the top of the computational box, since there the attributions are unambiguous. When orthogonality is re-established, so is the expected partition of the fast and Alfv\'en wave energy flux directions (see Section \ref{sec:flux} later). Irrespective of the assumed partition in the intervening region, the final fast and Alfv\'en fluxes at the top are definitive. The calculations presented here may therefore be regarded as scattering experiments: for a given incident fast wave at the bottom, how much energy emerges at the top and what form does it take?

%%%%
\subsection{Specific Model}  \label{model}

Consider a uniform magnetic field inclined at angle $\theta$ from the vertical in the $x$-$z$ plane and embedded in an atmosphere with density (and hence Alfv\'en speed) that is predominantly vertically stratified below $z\approx0$, becoming gradually more structured by field line above this level, representing a periodic ensemble of inclined ``flux tubes''.  A convenient form is
\begin{equation}
\frac{1}{a^2} = \frac{1}{a_0(z)^2} - \epsilon(z)\left[\cos(x-z\tan\theta)+\cos y\right],  \label{atm}
\end{equation}
where $a_0<1$ is monotonic increasing and $0\leq\epsilon(z)<\half$. The specific choice
\begin{equation}
a_0(z)^{2}=\frac{1+\delta+(1-\delta)\tanh(z/h_2)}{2(\rme^{-z/h_1}+1)},  \label{a0}
\end{equation}
with ``chromospheric" scale height $h_1>0$, and transition region thickness $h_2$. Note that $a_0^2\sim\delta\,\rme^{z/h_1}$ as $z\to-\infty$ and $a_0^2\to1$ as $z\to+\infty$. The parameter $\delta<1$ therefore characterizes the transition region density jump.

The periodicity $2\upi$ of the flux tubes in $x$ and $y$ defines the unit of length, and the normalization of Equation (\ref{atm}) defines the unit of time. This structure smoothly transitions from an exponentially decreasing behaviour with $z$ and density scale height $h_1$ in $z<0$ to a periodic horizontally structured tube ensemble in $z>0$. 

\begin{figure}
\begin{center}
\includegraphics[width=0.9\hsize]{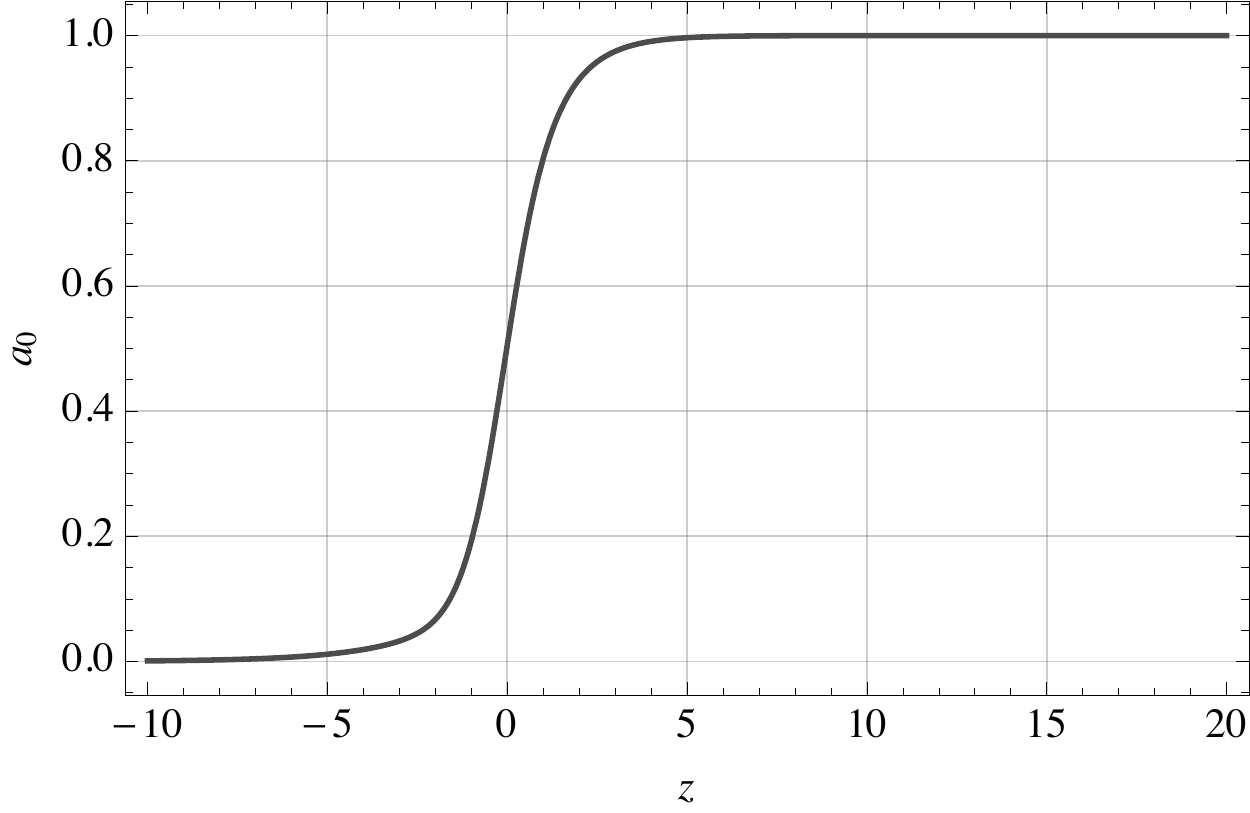}
\caption{The vertically stratified basic Alfv\'en speed $a_0$ as a function of height $z$ for the case $h_1=h_2=1$, $\delta=0.02$.}
\label{fig:a0}
\end{center}
\end{figure}

Reckoning on a tube periodicity of around 1 Mm (that is, 1 Mm scales to $2\upi$ in dimensionless units), reasonable parameter values for the solar chromospheric scale height and transition region thickness are $h_1=h_2=1$, with $\delta=0.02$. This case is illustrated in Figure \ref{fig:a0}.

The flux tubes are made to fade out over distance $W$ above $z=L$ by prescribing $\epsilon(z)=\half\epsilon_0\left(1-\tanh[(z-L)/W]\right)$. 

What should be expected from such a model? At large negative $z$, the atmosphere is effectively  plane parallel, with Alfv\'en speed increasing exponentially with height. This causes an upgoing fast wave to refract, and indeed reflect if its frequency $\omega$ is low enough that it does not reach $z\approx0$, where the atmosphere transitions to vertically unstratified. If the magnetic field is not vertical and if $k_y\neq0$, there is fast-to-Alfv\'en conversion associated with the vertical stratification (see \citealt{CalHan11aa} for the case of Alfv\'en speed increasing exponentially with height, and \citealt{HanCal12aa} for when there is a steep ``transition region'' as well). Additionally, the horizontal structuring that becomes apparent at $z\gtrsim 0$ will be responsible for the well-known damping of kink waves via Alfv\'en-continuum resonant absorption  \citep{RudRob02aa}.

%%%%%%%%%%%
\subsection{Fourier Formulation}
Expand
\begin{subequations}
\begin{align}
\Phi(\x,t) &= \sum_{m=-\infty}^\infty\sum_{n=-\infty}^\infty \phi_{mn}(z,t)\,\rme^{\rmi\left[(m+r)x+(n+s)y\right]},\\[4pt]
\Psi(\x,t) &= \sum_{m=-\infty}^\infty\sum_{n=-\infty}^\infty \psi_{mn}(z,t)\,\rme^{\rmi\left[(m+r)x+(n+s)y\right]},
\end{align}
\end{subequations}
where $r=m_0/M$ and $s=n_0/N$ are rational numbers (with $m_0$ and $M$ relatively prime, and similarly for $n_0$ and $N$) characterizing the $(2M\upi,\,2N\upi)$ periodicities of the initial or boundary conditions. Steady oscillations are sought, with $\exp(-\rmi\,\omega\,t)$ time dependence.

The various operations in Equation (\ref{PhiPsi}) are easily rendered in Fourier space.
For example, the Laplacian $\nabla^2$ is equivalent to $\upartial_z^2-(m+r)^2-(n+s)^2$. Similarly, $\upartial_y\equiv\rmi(n+s)$, $\upartial_\parallel\equiv\cos\theta\,\upartial_z+\rmi(m+r)\sin\theta$, and $\upartial_\perp\equiv\rmi(m+r)\cos\theta-\sin\theta\,\upartial_z$.

The effect of multiplying by $a^{-2}$, as given by Equation (\ref{atm}), is to scatter in $m$ and $n$ space by $\pm1$ in each direction. This couples the otherwise independent Fourier modes. 

The simple regular structure is of course an idealization. A more random placement may result in differences of detail, but would not be expected to change the basic behaviour if the tube separations are comparable. Furthermore, the ability to orient the incident wave more-or-less arbitrarily, via $k_x=r$ and $k_y=s$, makes the square structure less special.

It is convenient to define $Y=(\Phi,\Psi,P,V)^T$, where $P=\upartial_\perp\Phi$ and $V=\upartial_\perp\Psi$. Then equations (\ref{PhiPsi}) may be represented in matrix operator form as
\begin{equation}
M_{mn}Y_{mn} = 
R_n\,Y_{m+1\,n}+L_n\,Y_{m-1\,n}+U_{n+1}Y_{m\,n+1}+U_{n-1}Y_{m\,n-1}.          \label{MY}
\end{equation}
Here
\begin{equation}
M_{mn}=
\begin{pmatrix}
0 & -\rmi\,(n+s)\,\mathcal{A} & \mathcal{F} & 0 \\
\rmi\,(n+s)\,\mathcal{F} & 0 & 0 & \mathcal{A} \\
-\upartial_\perp & 0 & I & 0 \\
0 & -\upartial_\perp & 0 & I
\end{pmatrix},
\end{equation}
where $\mathcal{F}=\nabla^2+\omega^2/a_0^2$ is the (zero $\epsilon$) fast wave operator, $\mathcal{A}=\upartial_\parallel^2+\omega^2/a_0^2$ is the Alfv\'en operator, and $I$ is the identity. 

On the right hand side
\begin{equation} \label{U}
U_n =
\frac{\omega^2\epsilon}{2}\begin{pmatrix}
0 & -\rmi(n+s)  &I & 0 \\
\rmi(n+s)  & 0 & 0 & I \\
0 & 0 & 0 & 0 \\
0 & 0 & 0 & 0
\end{pmatrix},
\end{equation}
$R_n=\rme^{\rmi\,z\tan\theta} U_n$, and $L_n=\rme^{-\rmi\,z\tan\theta}U_n$.

In practice, the equations are solved in finite difference form, with $N$ grid points $z=(z_1,\ldots,z_N)^T$ vertically. Thus each of the entries in the above matrices become $N\times N$ submatrices, with $M_{mn}$, $R_n$, $L_n$, and $U_n$ becoming $4N\times4N$. Off-diagonal (band) terms in the submatrices of $M$ are generated by $z$-derivatives in $\mathcal{F}$, $\mathcal{A}$ and $\upartial_\perp$. The width of the bands depends on the order of the finite difference derivatives employed (arbitrary order is coded, with $12^\text{th}$ order generally used).

The various functions of $z$ appearing in these matrices (specifically $a_0^2$, $\epsilon$, and $\rme^{\pm\rmi\,z\tan\theta}$) manifest as diagonal matrices. Thus for example, $\epsilon(z)$ becomes $\diag[\epsilon(z_1),\ldots,\epsilon(z_N)]$ in each $N\times N$ submatrix.

Boundary conditions are required to complete the specification of the problem.

%%%
\subsubsection{Boundary Conditions}  \label{BCs}
The top and bottom of the computational region, $z_\text{top}$ and $z_\text{bot}$, are placed in regions where the horizontal structuring is negligible: $z_\text{top}-L\gg W$ and $(-z_\text{bot})\gg h$. At these extremes, it is imposed that there are no incoming waves, save for the $m=n=0$ fast wave at $z_\text{bot}$.

The dispersion relation resulting from Equations (\ref{PhiPsi}) is
\begin{equation}
(\kperp^2+k_y^2)(\omega^2-a^2k^2)(\omega^2-a^2\kpar^2)=0,
\end{equation}
where $k^2=|\bk|^2=k_x^2+k_y^2+k_z^2=\kperp^2+k_y^2+\kpar^2$.

The pure fast wave and Alfv\'en wave dispersion relations, $\omega^2=a^2k^2$ and $\omega^2=a^2k_\parallel^2$ respectively, may be solved for the $z$-component of $\bk$. For the fast wave
\begin{subequations}
\begin{align}
&k_z = \pm\sqrt{\frac{\omega^2}{a^2}-(m+r)^2-(n+s)^2}= \pm K_{mn} \\[4pt]
&\text{with eigenvector } 
\begin{pmatrix}
\Phi\\ \Psi
\end{pmatrix}
=
\begin{pmatrix}
1 \\ 0
\end{pmatrix}.
\end{align}
For the Alfv\'en wave
\end{subequations}
\begin{subequations}
\begin{align}
&k_z = \pm\frac{\omega}{a}\sec\theta-(m+r)\tan\theta =\kappa_{mn}^\pm  \\[4pt]
&\text{with eigenvector } 
\begin{pmatrix}
\Phi\\ \Psi
\end{pmatrix}
=
\begin{pmatrix}
0 \\ 1
\end{pmatrix}.
\end{align}
\end{subequations}
In each case, the positive root is upgoing. 

There is though another spurious ``mode'', independent of frequency, with dispersion relation $\kperp^2+k_y^2=0$, i.e., $\kperp=\mp\rmi\,k_y$, or 
\begin{subequations}
\begin{align}
&k_z =(m+r)\cot\theta\pm\rmi\,(n+s)\csc\theta=\sigma_{mn}^\pm, \\[4pt]
&\text{with eigenvector } 
\begin{pmatrix}
\Phi\\ \Psi
\end{pmatrix}
=
\begin{pmatrix}
1 \\ \mp\rmi
\end{pmatrix}.
\end{align}
\end{subequations}
Note that $\exp[\rmi \sigma_{mn}^\pm z]$ are the two linearly independent solutions of $\nabla_\text{p}^2\Upsilon=0$. These solutions represent the gauge freedom discussed in Section \ref{fund}.

The spurious mode results from the system being $6^\text{th}$ order (for each $m,n$ pair). It needs to be removed by appropriate application of auxiliary conditions. In light of its eigenvectors, this may be accomplished by using
\begin{equation}
\Lambda_{mn}=\Phi_{mn}''+K_{mn}^2\Phi_{mn} \label{Lambda}
\end{equation}
to filter out the fast wave, and then applying (optionally) zero or evanescence conditions at the boundaries.
This prescription works by removing the pure oscillatory fast mode $\exp(\pm\rmi\,K_{mn}z)$, leaving an arbitrary linear combination of $\exp(\rmi\,\sigma_{mn}^\pm z)$.

Based on these behaviours, the six boundary conditions applied are,
\begin{subequations}
\label{BC}
\renewcommand{\theequation}{\theparentequation.\Roman{equation}}
\begin{gather}
\noalign{\noindent at $z_{\text{bot}}$:}
\Lambda_{mn}=0 \quad\text{ or }\quad  \Lambda_{mn}'-\ri\, \sigma_{mn}^\mp \Lambda_{mn}=0 \label{I}\\[2pt]
\chi_{mn}'+\rmi\,K_{mn}\chi_{mn}=-k_{\text{p}}^2\, \delta_{m0}\delta_{n0}  \label{II}\\[2pt]
\zeta_{mn}'-\rmi\,\kappa^-_{mn}\zeta_{mn}=0;  \label{III}\\[4pt]
\noalign{\noindent at $z_{\text{top}}$:}
\Lambda_{mn}=0  \quad\text{ or }\quad  \Lambda_{mn}'-\ri\, \sigma_{mn}^\pm \Lambda_{mn}=0 \label{IV}\\[2pt]
\chi_{mn}'-\rmi\,K_{mn}\chi_{mn}=0  \label{V}\\[2pt]
\zeta_{mn}'-\rmi\,\kappa^+_{mn}\zeta_{mn}=0.  \label{VI}
\end{gather}
\end{subequations}
The upper sign is taken in Equations (\ref{I}) and (\ref{IV}) if $(n+s)\csc\theta>0$, and the lower sign otherwise.

The radiation conditions (\ref{II}), (\ref{III}), (\ref{V}) and (\ref{VI}) are applied to $\chi$ and $\zeta$ rather than the potentials, as they  unambiguously represent the fast and Alfv\'en waves respectively. The solutions obtained are therefore robust irrespective of gauge,\footnote{In that regard, an arbitrary gauge could be applied to the solution as a post-process without invalidating it.} though their attributions via the potentials to mode type depend on gauge choice. The prescriptions (\ref{I}) and (\ref{IV}) perform well in that regard, subject to the accuracy of the WKB approximation, with little practical difference between the solutions found with zero or evanescence boundary conditions. The former are therefore used for simplicity. 

The six boundary conditions (\ref{BC}) are inserted into the $4N\times4N$ matrix equation (\ref{MY}), replacing respectively rows $1$, $N+1$, $3N+1$, $N$, $2N$, and $4N$, i.e.,~the top and bottom $\Phi$, $\Psi$, and $V$ equations.\footnote{This choice was arrived at by experiment. Solutions obtained with it coincide very well with shooting method solutions, which do not need to make such substitutions.} The injection equation (\ref{II}) contributes an inhomogeneous source term $S=(0,0,\ldots,0,-k_\text{p}^2,0,\ldots,0)^T$, where the sole nonzero term is at entry $N+1$. The full equation then takes the form
\begin{multline}
M_{mn}Y_{mn} 
-R_n\,Y_{m+1\,n}-L_n\,Y_{m-1\,n}-U_{n+1}Y_{m\,n+1}-U_{n-1}Y_{m\,n-1}\\
 = S_{mn},  \qquad      \label{MBY}
\end{multline}
where it is to be understood that the six boundary replacements in $M_{mn}$ etc.~have been made, as described.

%%%%%%
\subsection{Wave Energy and Flux}\label{sec:flux}
The quadratic wave energy equation in conservation form may be constructed directly from Eqn (\ref{basiceqn}) by contraction with $\boldv=\upartial_t\bxi$. After some algebra, it follows that
\begin{equation}
\upartial_tE+\Div\F=0,
\end{equation}
where
\begin{equation}
E=\half\rho_0|\boldv|^2+\frac{B_0^2}{2\umu}\left(\chi^2+|\upartial_\parallel\bxi|^2\right)
\end{equation}
is the wave energy density and 
\begin{equation}
\F=-\frac{B_0^2}{\umu}\left(\chi\boldv+(\boldv\vdot\upartial_\parallel\bxi)\e_\parallel\right)  \label{flux}
\end{equation}
is the wave energy flux. The field-directed unit vector is denoted by $\e_\parallel$.
The perpendicular-to-the-field term in $\F$ proportional to $\chi\boldv$ is just the rate of working of the fast wave's magnetic pressure perturbation. The field-aligned tension-related term $(\boldv\vdot\upartial_\parallel\bxi)\e_\parallel$ on the other hand includes both fast and Alfv\'en contributions.

In terms of the complex solutions of Equations (\ref{PhiPsi}),  the vertical component to wave-energy flux associated with Fourier numbers $m$ and $n$ and averaged over a period in both $x$ and $y$ is
\begin{equation}
F_{mn}=F_0\im\left\{\chi\xi_z^*+(\bxi^*\vdot\upartial_\parallel\bxi)\cos\theta\right\},  \label{fluxz}
\end{equation}
where $F_0=\omega B_0^2/\umu$. With the amplitude of the incident fast potential set to $|\Phi_{00}^{\text{inc}}|=1/(2|k_z|)$ from boundary condition (\ref{II}), the incident flux may be normalized to unity by setting
\begin{equation}
F_0=\frac{4|k_z|^2}{\re\{k_z\}\,k_{\text{p}}^2}\Biggr|_{z=z_{\text{bot}}}.
\end{equation}
The fast wave vertical wavenumber $k_z=K_{00}$ is evaluated at the base, with $k_x=r$, $k_y=s$, and $k_{\text{p}}^2=k_\perp^2+k_y^2=(k_x\cos\theta-k_z\sin\theta)^2+k_y^2$. It is assumed that $K_{00}$ is real, i.e., that the fast wave is travelling, not evanescent, at the injection height.

Total $x$-$y$-averaged flux $\sum_m\sum_n F_{mn}$ is independent of height, but individual components $F_{mn}$ are not in general, since energy cascades in Fourier space.

\begin{figure*}
\begin{center}
\hfill\includegraphics[width=0.4\textwidth]{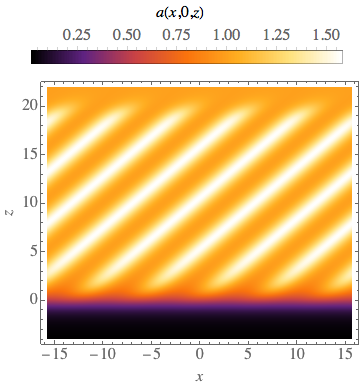} \hfill
\includegraphics[width=0.4\textwidth]{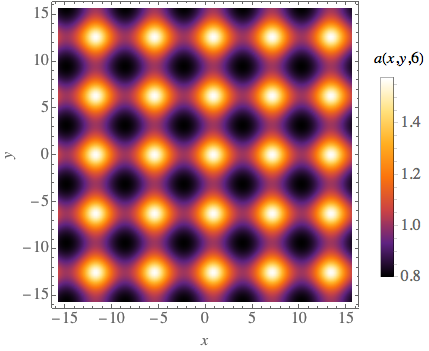}\hfill
\caption{Representative image of the flux tubes, as represented by the Alfv\'en speed $a(x,y,z)$, in the vertical $x$-$z$ plane at $y=0$ (left) and the horizontal $x$-$y$ plane at $z=6$ (right) for the case $\theta=50^\circ$, $\epsilon_0=0.3$, $h_1=h_2=1$, $\delta=0.02$, $L=20$, $W=1$. With $M=N=5$ (corresponding to $r=1/5$, $s=4/5$ for example), the $x$-$y$ plot represents the periodicity in both directions.}
\label{fig:tubes}
\end{center}
\end{figure*}

%%%%%%%%%%%%%%%%%%%%%%%%%%%%%%%%%%%%%%%%%%%%%%%%
\section{Numerical Solution}
Energy is injected into the system through $z_{\text{bot}}$ in mode $m=n=0$ only. It is distributed between modes via the couplings to its North, South, East and West neighbours in Fourier space, i.e., from $(0,0)$ to $(0,\pm 1)$ and $(\pm 1,0)$, and thence on to all other modes. Since steady oscillations are sought ($\omega$ real), total flux $\sum_m\sum_n F_{mn}$ summed over all modes remains independent of $z$.  

Strictly, Equations (\ref{MBY}) should be solved simultaneously for all modes $(m,n)$, but the computational expense is prohibitive. 

Several solution strategies present themselves. Block-Jacobi iteration
\begin{multline}
M_{mn}Y_{mn}^{(j+1)} =
R_n\,Y_{m+1\,n}^{(j)}+L_n\,Y_{m-1\,n}^{(j)}+U_{n+1}Y_{m\,n+1}^{(j)}+U_{n-1}Y_{m\,n-1}^{(j)}\\
+S_{mn},     \qquad \qquad   \label{Jacobi}
\end{multline}
is cheap, even with several thousand grid points in $z$, since $M_{mn}$ may be LU-decomposed or similar once for each $(m,n)$ pair, resulting in each iteration requiring only inexpensive back-substitution. It is also readily parallelized. This scheme does not strictly conserve flux, but once converged does so to the required tolerance. Convergence is contingent on the appropriate spectral radius being less than unity. The spectral radius is not determined explicitly, but experience indicates that the scheme is convergent in some cases and divergent in others. It is certainly convergent for small enough $\epsilon_0$.

Convergence may be improved by taking advantage of block-tridiagonal structure generated by adopting an implicit formulation in one direction. For example, implicit coupling in the horizontal ($m$) direction yields
\begin{multline}
M_{mn}Y_{mn}^{(j+1)} -R_n\,Y_{m+1\,n}^{(j+1)}-L_n\,Y_{m-1\,n}^{(j+1)}=U_{n+1}Y_{m\,n+1}^{(j)}+U_{n-1}Y_{m\,n-1}^{(j)}\\
+S_{mn}.    \qquad \qquad   \label{himplicit}
\end{multline}
A vertical formulation is defined analogously:
\begin{multline}
M_{mn}Y_{mn}^{(j+1)} -U_{n+1}Y_{m\,n+1}^{(j+1)}-U_{n-1}Y_{m\,n-1}^{(j+1)}=R_n\,Y_{m+1\,n}^{(j)}+L_n\,Y_{m-1\,n}^{(j)}\\
+S_{mn}.    \qquad \qquad   \label{vimplicit}
\end{multline}
Both of these schemes conserve flux along their implicit direction, but not perpendicularly. Again, convergence  redresess this. Experience suggests that these line-implicit schemes converge very rapidly, though the decompositions of the left hand sides are very expensive and memory-intensive. Once calculated though, it is comparatively cheap to be apply them recursively till convergence. Nevertheless, simple Jacobi iteration is far less memory-intensive and far quicker, provided it converges.

The adopted finite difference scheme in $z$ typically uses $12^\text{th}$ order finite differences on a regular grid of about 8000--12000 points. The code though is written to accommodate arbitrary order and an optionally stretched grid. The various Jacobi or line-iterations are performed in parallel on multi-core machines. Iteration is continued until
\begin{equation}
\max_{m,n,z} \left| Y_{mn}^{(j+1)}-Y_{mn}^{(j)} \right| < 10^{-4}.  \label{converge}
\end{equation}
Memory is the determining limitation in the line-implicit methods.

%%%%%%%%%%%%%%%%%%%%%%%%%%%%%%%%%%%%%%%%%%%%%%%%
\section{Results}
The coupled ordinary differential equations (\ref{MBY}) are solved numerically in an atmosphere structured as displayed in Figure \ref{fig:tubes}.

%%%%%%%%%%%%%%%%%%%
\subsection{No Tubes; $\epsilon=0$}
For purposes of comparison, it is of interest to first solve the problem for the case $\epsilon_0=0$, where there are no ``flux tubes''. Only $m=n=0$ need be considered, as there is no coupling to the other Fourier modes. This is the case extensively explored by \citet{CalHan11aa}, though with a different Alfv\'en speed profile.

\begin{figure}
\begin{center}
\includegraphics[width=.9\hsize]{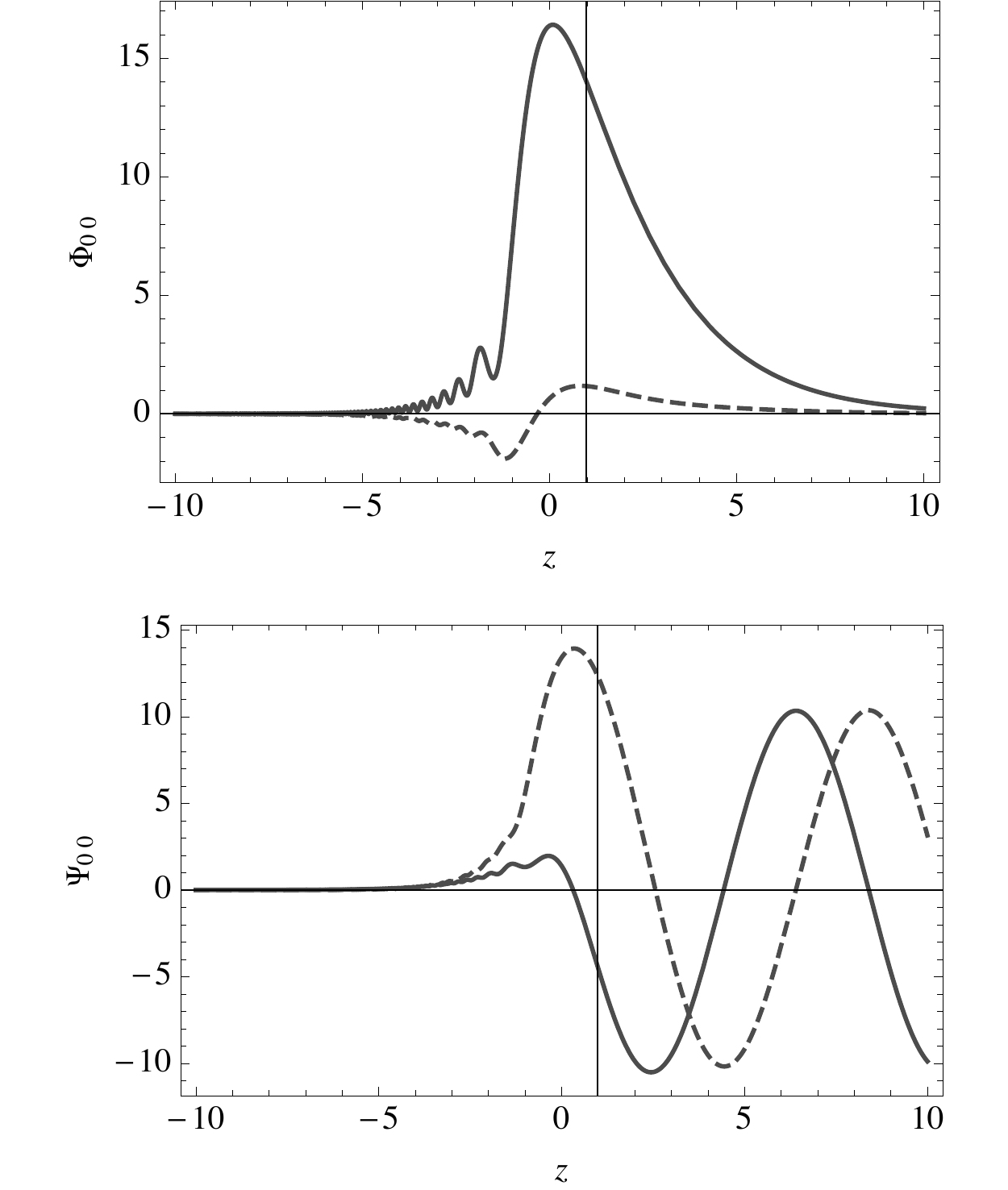}
\caption{$\Phi_{00}$ and $\Psi_{00}$ as functions of height $z$ for the case of Fig.~\ref{fig:tubes} with $r=1/5$, $s=4/5$, $\omega=0.8\,\omega_0$ (reflecting fast wave), and $\epsilon_0=0$ (no tubes). The evanescence of the fast wave ($\Phi$) and the upward travelling nature of the Alfv\'en wave ($\Psi$) are apparent. The vertical line indicates the position of the fast wave reflection point. Real and imaginary parts are shown as full and dashed curves respectively.}
\label{fig:PhiPsi000}
\end{center}
\end{figure}

\begin{figure}
\begin{center}
\includegraphics[width=.9\hsize]{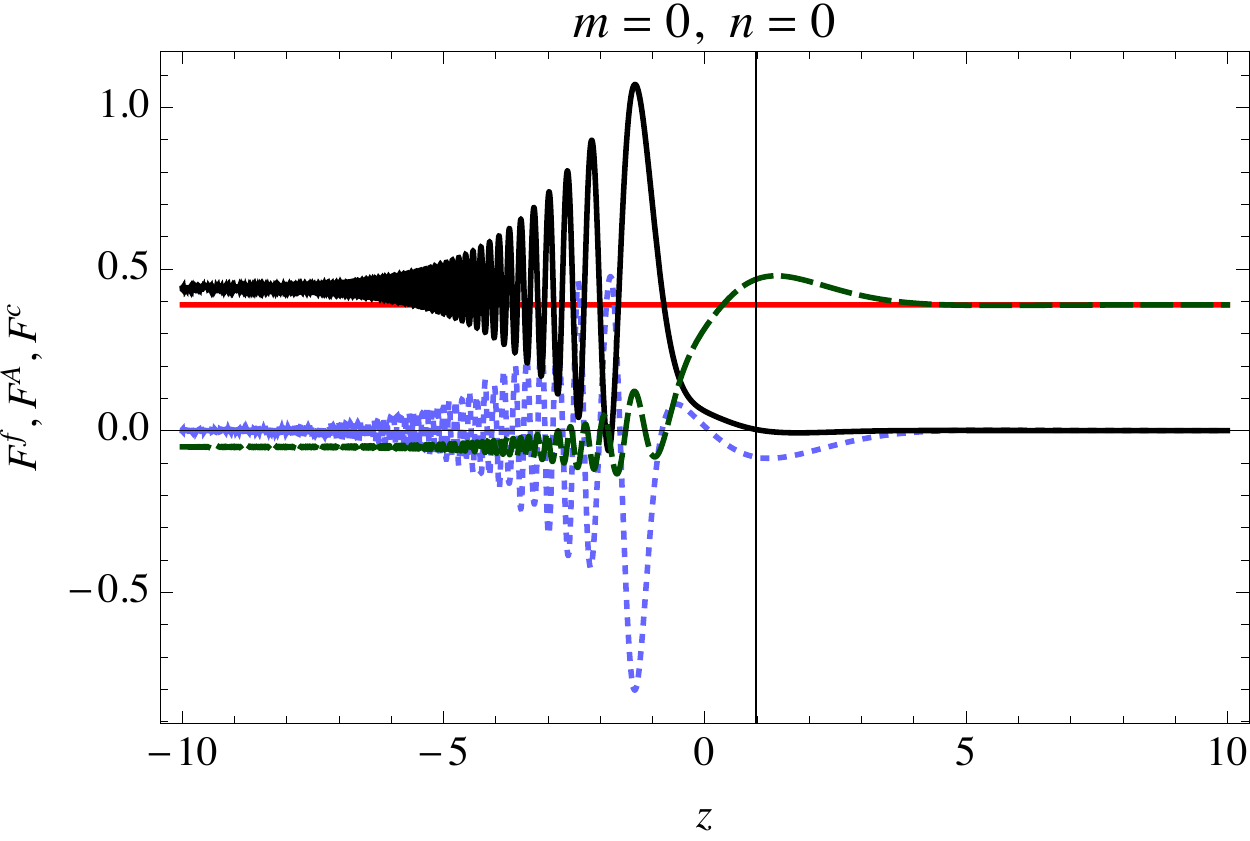}
\caption{The fast (full black), Alfv\'en (dashed), and cross (dotted) fluxes against $z$ for the case of Fig.~\ref{fig:PhiPsi000}. The total flux $F=F^{\text{f}}+F^{\text{A}}+F^{\text{c}}=0.39$ (horizontal red line) is constant, as required since there is no lateral energy loss.}
\label{fig:F000}
\end{center}
\end{figure}

The maximum Alfv\'en speed attained is 1, so any $m=n=0$ fast wave with frequency $\omega$ will reflect if $\omega<\omega_0=(r^2+s^2)^{1/2}$. If $\omega\ge\omega_0$ there is no (total) reflection, and the fast wave propagates (partially) to infinity. It is convenient to define the parameter $\alpha$ such that $\omega=\alpha\, \omega_0$, so the incident wave reflects if $\alpha<1$. 

For the first case, the incident fast wave has frequency parameter $\alpha=0.85$, so it is trapped.
Figures \ref{fig:PhiPsi000} and \ref{fig:F000} show the coupling effect of nonzero $k_y=s$, as anticipated from Equations (\ref{PhiPsi}). The incident fast mode (unit flux) from below mostly reflects, but with 39\% escaping the top as an Alfv\'en wave. There is no fast wave flux at the top, as expected. Outgoing flux at the bottom consists of 56\% in the fast wave and 5\% in the Alfv\'en wave.

The cross-flux, shown dotted in Figure \ref{fig:F000}, consists of those terms in the quadratic formula (\ref{fig:F000}) containing both $\Phi$ and $\Psi$. It is therefore a direct measure of the fast/Alfv\'en coupling. Once it dies out with increasing $z$, the decoupling of the fast and Alfv\'en components is essentially complete.

\begin{figure}
\begin{center}
\includegraphics[width=.9\hsize]{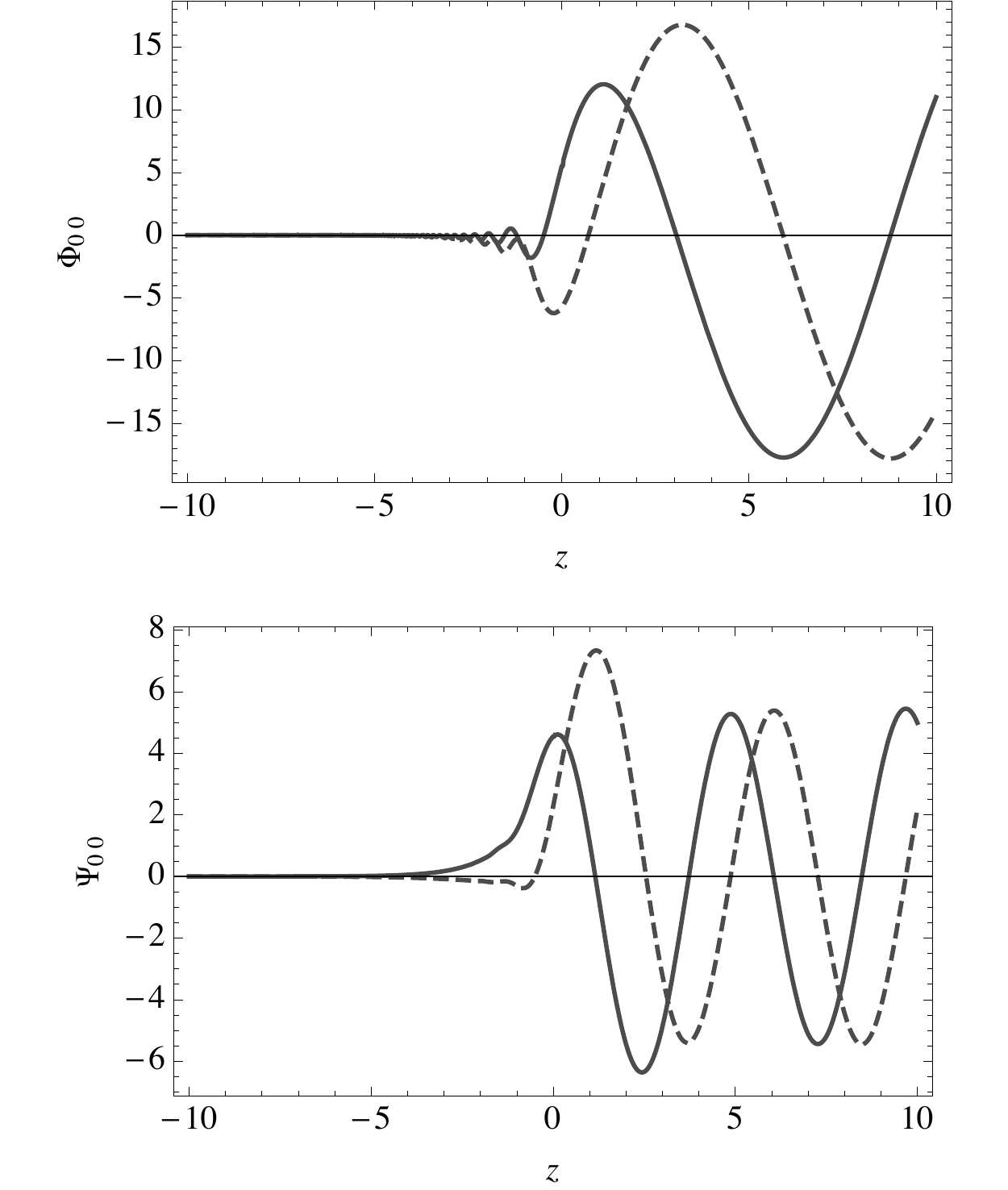}
\caption{$\Phi_{00}$ and $\Psi_{00}$ as functions of height $z$ for the case of Fig.~\ref{fig:tubes} with $r=1/5$, $s=4/5$, $\omega=1.2\,\omega_0$ (transmitting fast wave), and $\epsilon_0=0$ (no tubes). The travelling natures of the fast wave ($\Phi$) and Alfv\'en wave ($\Psi$) is apparent.}
\label{fig:PhiPsi000u}
\end{center}
\end{figure}

\begin{figure}
\begin{center}
\includegraphics[width=.9\hsize]{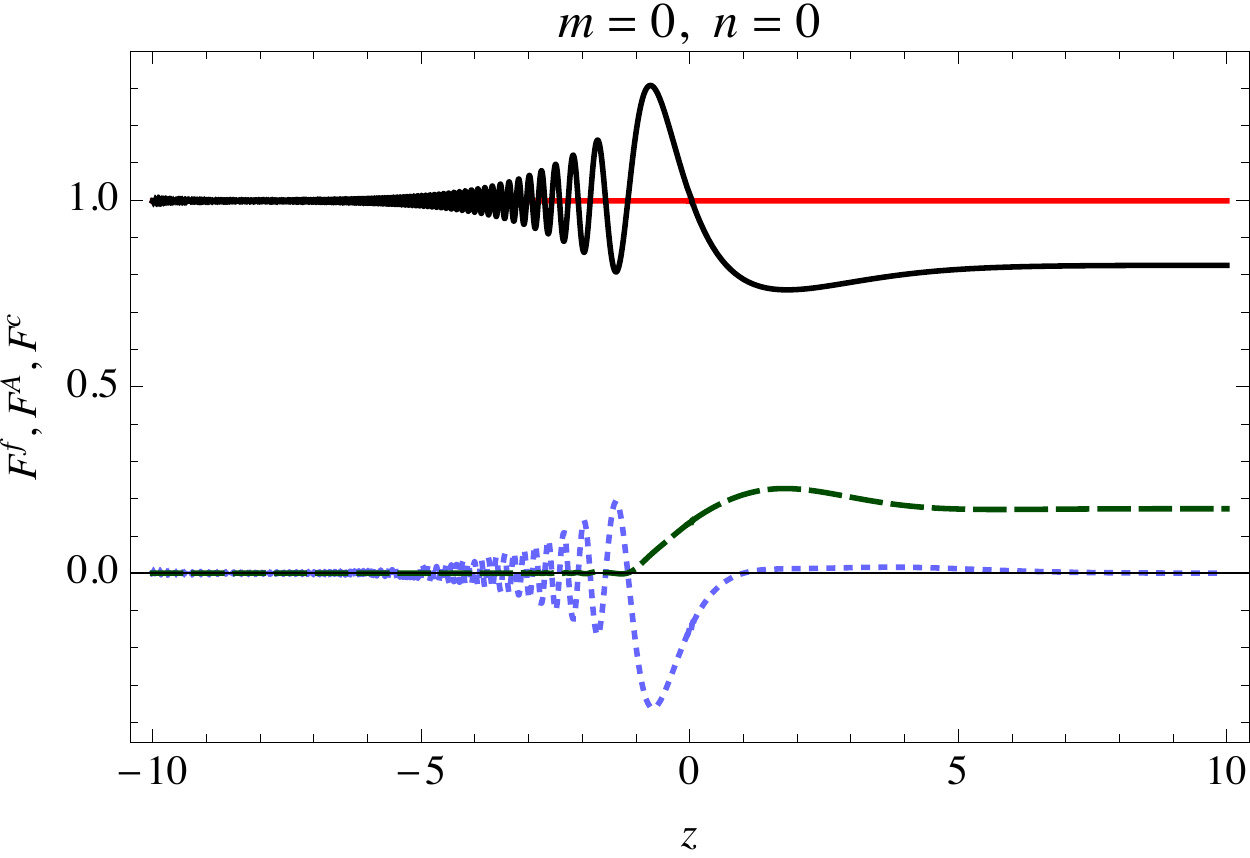}
\caption{The fast (full black), Alfv\'en (dashed), and cross (dotted) fluxes against $z$ for the case of Fig.~\ref{fig:PhiPsi000}. The total flux $F=F^{\text{f}}+F^{\text{A}}+F^{\text{c}}=0.998$ is plotted in red.}
\label{fig:F000u}
\end{center}
\end{figure}

For the second case, $\alpha=1.2$ is chosen. In this circumstance, 83\% of the injected flux escapes at the top as a fast wave, and 17\% as an Alfv\'en wave. Less than 0.3\% escapes at the bottom as a reflected fast wave. Figures \ref{fig:PhiPsi000u} and \ref{fig:F000u} illustrate the nature of this case.

The cursory discussion of the $\epsilon=0$ case presented here is for purposes of comparison. Atmospheres with periodic tube structures will be the subject of investigation from now on.

%%%%%%%%%%%%%%%%%%%
\subsection{Tubes; $0<\epsilon<0.5$}
The situation of prime interest is where the frequency of the incident wave is such that fast waves are trapped in low Alfv\'en speed (high density) tubes. This requires $1-\epsilon_0 < \alpha < 1$ (Section \ref{tubes}). The case $1<\alpha<1+\epsilon_0$ in which these tubes merge into a contiguous ``Swiss cheese'' is also briefly addressed (Section \ref{cheese}).

%%%%%
\subsubsection{$\alpha=0.85$} \label{tubes}
Restricting the frequency parameter to the lower half of the band $(1-\epsilon_0,\,1+\epsilon_0)$ restricts the $x$-$y$ surface area where the fast wave may propagate to discrete tubes, as commonly envisaged for kink waves. For example, with $\epsilon_0=0.3$ and $\alpha=0.85$, those tubes occupy 31\% of the area. With $\alpha=1$ it is 50\%, but the ``tubes'' become contiguous. For the moment, only the discrete-tube case $\alpha<1$ is examined.

A representative case $r=1/11$, $s=3/5$, $\theta=50^\circ$, $\epsilon_0=0.3$, $\alpha=0.85$ , $L=12$, $W=1$, $z_\text{bot}=-8$, $z_\text{top}=14$ with $-3\le m,\,n\le3$ is chosen by way of illustration. It is sufficient to illustrate the general features of the system. Jacobi iteration is used to obtain the converged solution.

\begin{figure}
\begin{center}
\includegraphics[width=\hsize]{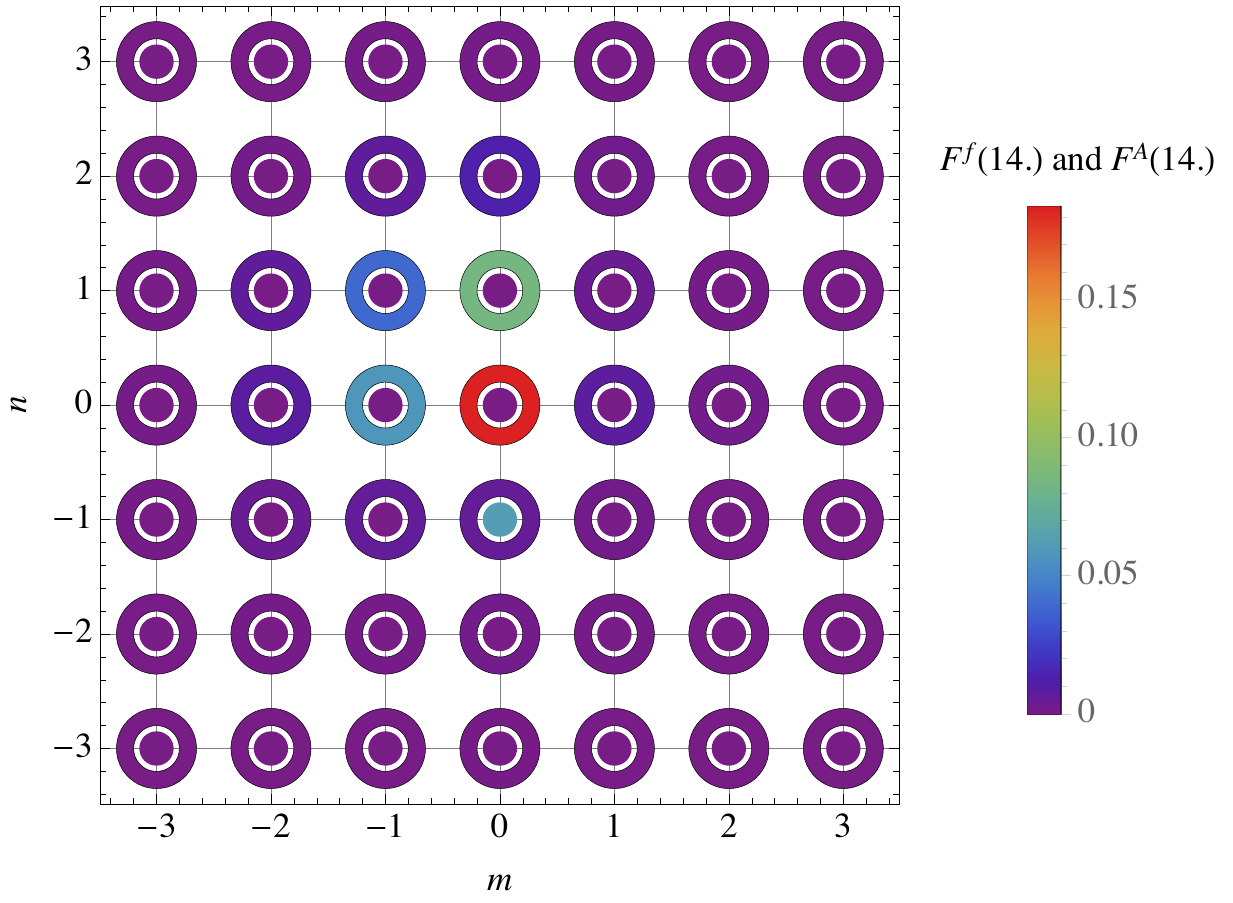}
\caption{Top Alfv\'en (outer annulus) and fast (inner disk) fluxes in the Fourier modes $m$, $n$ for the case $r=1/11$, $s=3/5$, $\theta=50^\circ$, $\epsilon_0=0.3$, $\alpha=0.85$ , $L=12$, $W=1$, $z_\text{bot}=-8$, $z_\text{top}=14$ with $-3\le m,\,n\le3$.}
\label{fig:kase4spec}
\end{center}
\end{figure}

Figure \ref{fig:kase4spec} shows the Alfv\'en and fast wave fluxes at the top of the box for the forty-nine central Fourier modes $-3\le m,\,n\le3$. It is clear that energy has not propagated very far in mode number by this level ($z_\text{top}=14$). This verifies that enough Fourier modes have been used. Larger mode sets are easily handled with the Jacobi process, but are computationally more problematic for line-implicit calculations. Increasing the number of modes beyond the current level does not seem to adversely affect stability for the cases examined.

The highest flux is Alfv\'enic for $m=n=0$, the only mode at which energy is injected (as a fast wave at $z_\text{bot}=-8$). However, there is also significant energy flux in the immediately surrounding modes, mostly Alfv\'enic, but strongly fast for the sole case $m=0$, $n=-1$. 

The reason for this is that the horizontal wavenumber $k_h=\sqrt{(m+r)^2+(n+s)^2}$ is a minimum there ($k_h=0.41$), making the $m=0$, $n=-1$ mode vertically propagating not only at the tube centres ($\alpha\omega_0\sqrt{1+\epsilon_0}=0.59$), but even (marginally) at the anti-tube centres ($\alpha\omega_0\sqrt{1-\epsilon_0}=0.43$). Consequently, for this case, the coronal $m=0$, $n=-1$ fast wave is not restricted to the low Alfv\'en speed tubes but is space-filling. All others Fourier modes are evanescent throughout, and so carry no fast wave energy vertically. Different choices of parameters may yield some modes for which $\alpha\omega_0\sqrt{1-\epsilon_0} < k_h < \alpha\omega_0\sqrt{1+\epsilon_0}$; these will be true propagating tube waves.

This illustrates how the scattering in Fourier space can open up one or more channels for fast mode propagation despite the original incident fast wave being evanescent. However, the Alfv\'en wave propagates in all channels; it is simply a matter of how much scatters into them from the incident wave.

\begin{figure*}
\begin{center}
\includegraphics[width=.9\hsize]{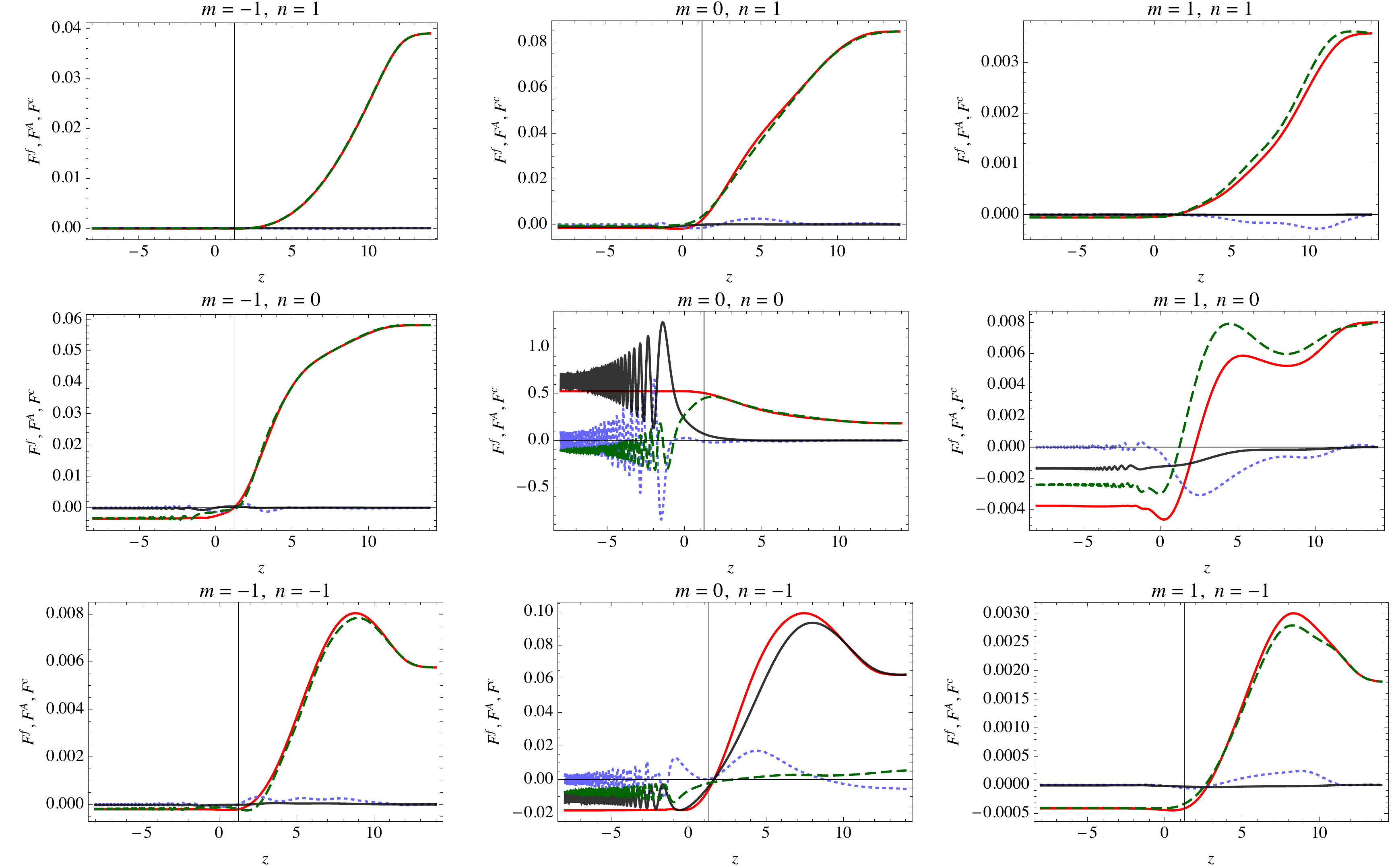}
\caption{Alfv\'en (green dashed), fast (black full), cross (dotted blue), and total (red full) fluxes in the Fourier modes $-1\le m,\,n\le1$ (labelled) as functions of height $z$ for the case of Figure \ref{fig:kase4spec}. Note the different flux scales in each panel.}
\label{fig:kase4Fmn}
\end{center}
\end{figure*}

Figure \ref{fig:kase4Fmn} shows these fluxes in detail as functions of height for $-1\le m,\,n\le1$. It is apparent that the injected fast flux in the central mode, $m=n=0$, quickly converts to Alfv\'en flux near the reflection height, and then that that Alfv\'en flux slowly decays with height as it is transferred to surrounding Fourier modes by phase mixing generated by the tube structure.

For the most part, the lost energy reappears as Alfv\'en flux in the surrounding modes, increasing in magnitude with increasing $z$, and levelling off only where the tubes fade out around $z=12$. However, the propagating fast mode in $m=0$, $n=-1$, is very striking. It grows rapidly over $0\lesssim z\lesssim7$, before itself becoming subject to phase mixing decay. It again levels off as the tubes fade out, but in a more realistic model with much longer tubes, would decay away almost completely, leaving only Alfv\'en energy in the system. (Jacobi iterations do not converge if the tubes are much longer, so they are restricted here for computational convenience.)

\begin{figure}
\begin{center}
\includegraphics[width=.9\hsize]{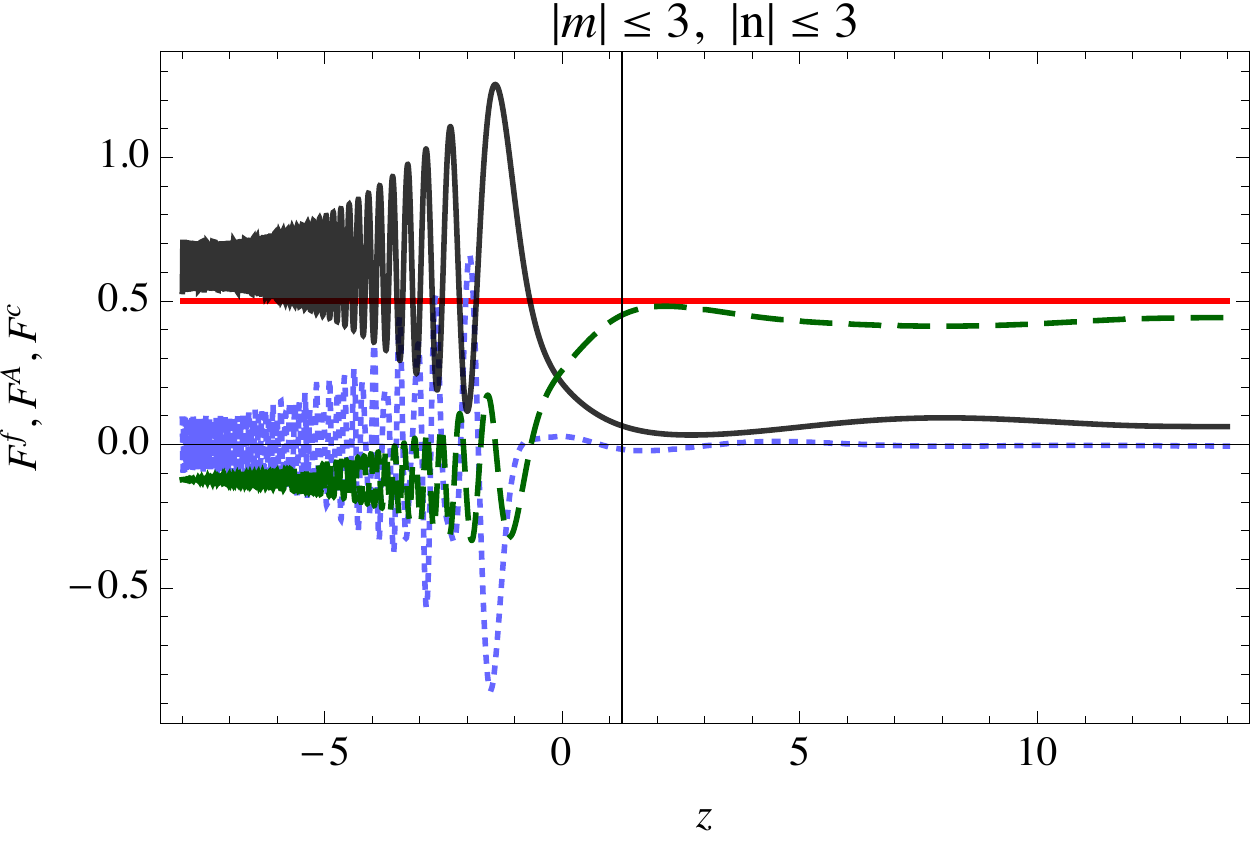}
\caption{Alfv\'en (green dashed), fast (black full), cross (dotted blue), and total (red full) fluxes summed over all Fourier modes $-3\le m,\,n\le3$ as functions of height $z$ for the case of Figure \ref{fig:kase4spec}.}
\label{fig:kase4Ftot}
\end{center}
\end{figure}

The total flux, summed over all modes, is depicted in Figure \ref{fig:kase4Ftot}. Two points to note are that the total flux is indeed independent of height (this is a sign of convergence of the Jacobi iterations), and that the overall flux is predominantly Alfv\'enic.

\begin{figure*}
\begin{center}
\includegraphics[width=.9\hsize]{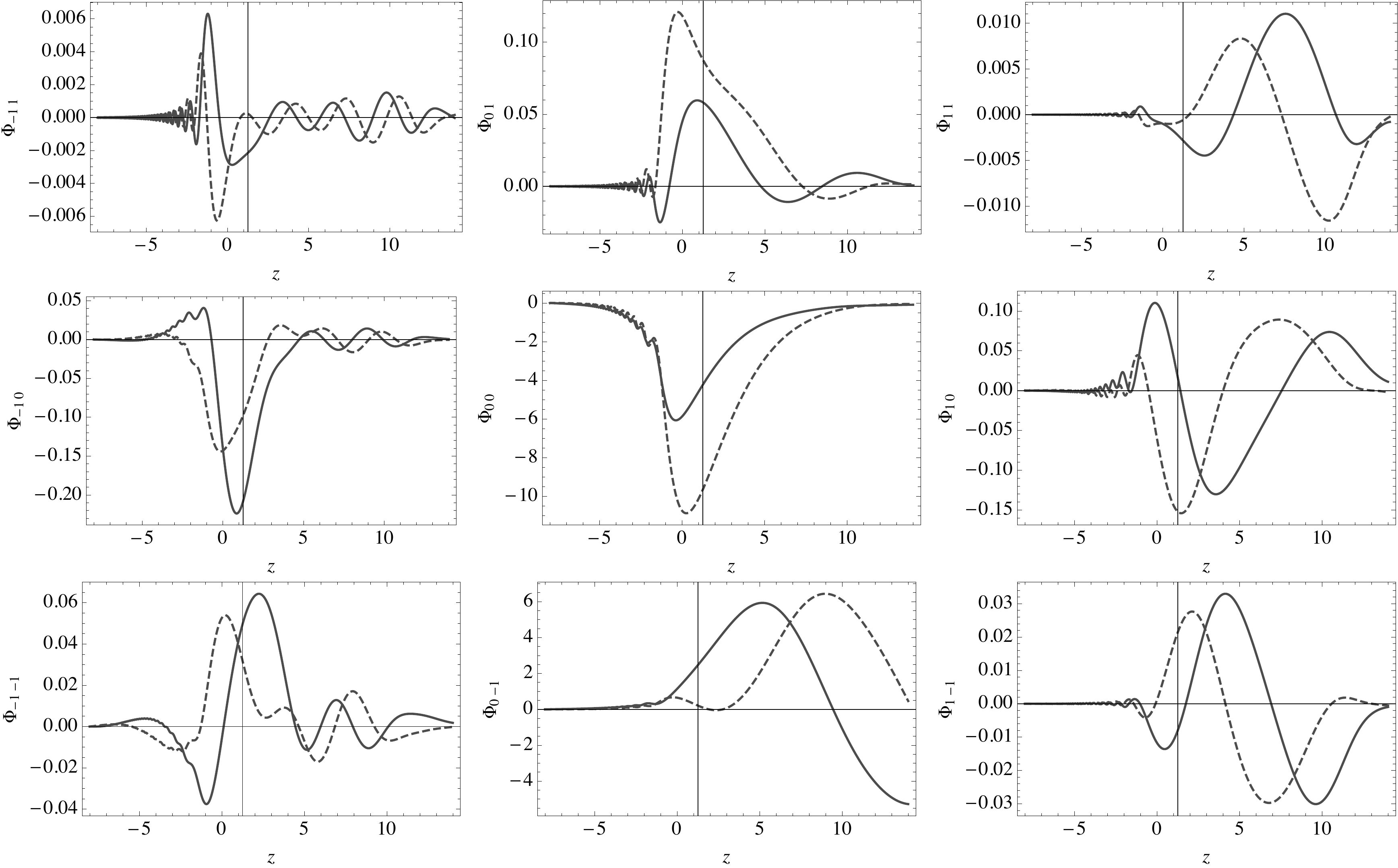}
\caption{Fast wave potential Fourier coefficients $\Phi_{mn}$ for $-1\le m,\,n\le1$ as functions of height for the case of Figure \ref{fig:kase4spec}. The modes are labelled as subscripts on the axis label ``$\Phi$''. Specifically, $m=-1,\,0,\,1$ left to right, $n=-1,\,0,\,1$ bottom to top.}
\label{fig:kase4Phimn}
\end{center}
\end{figure*}

\begin{figure*}
\begin{center}
\includegraphics[width=.9\hsize]{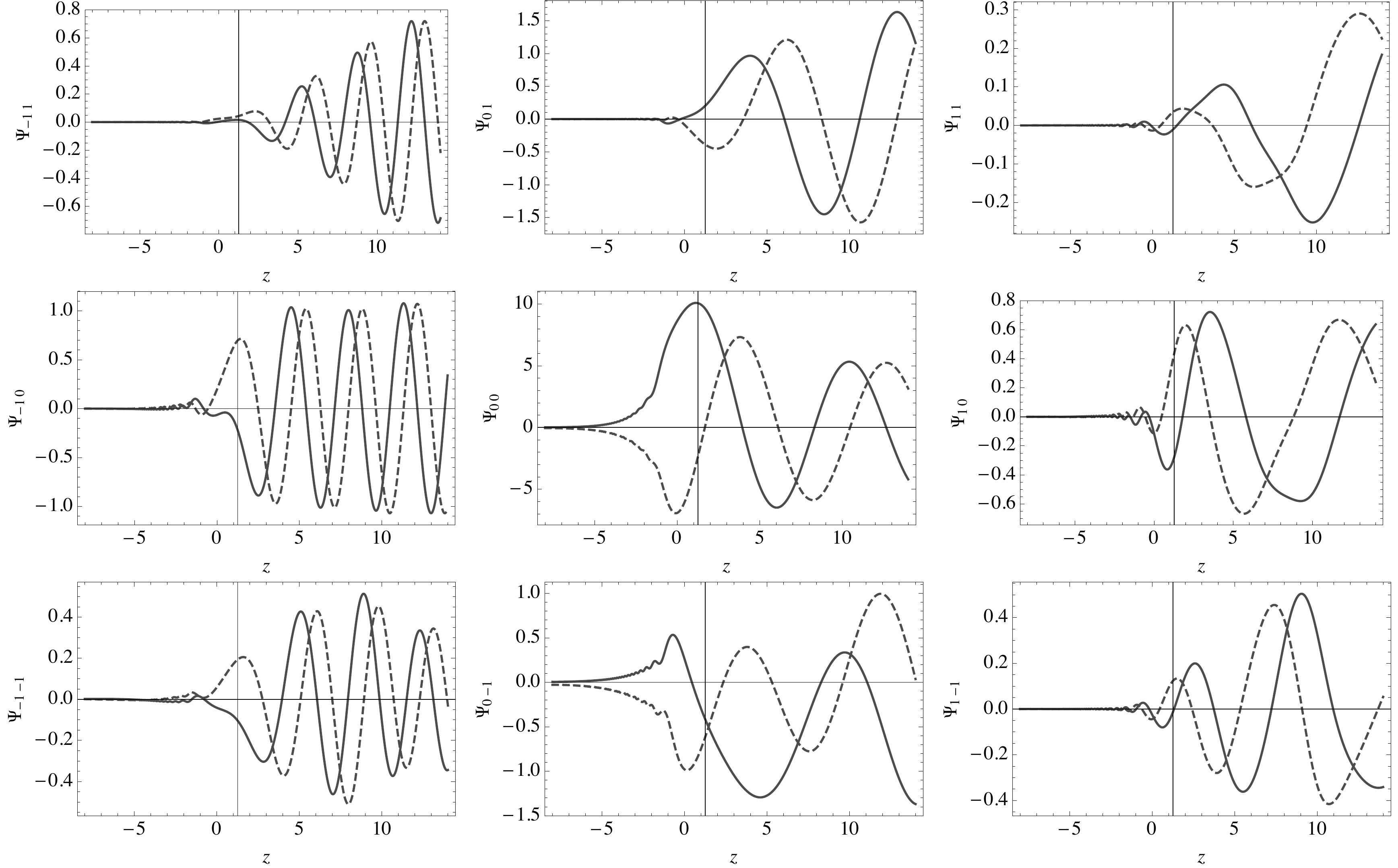}
\caption{Alfv\'en wave potential Fourier coefficients $\Psi_{mn}$ for $-1\le m,\,n\le1$ as functions of height for the case of Figure \ref{fig:kase4spec}. }
\label{fig:kase4Psimn}
\end{center}
\end{figure*}

Figures \ref{fig:kase4Phimn} and \ref{fig:kase4Psimn} display the central nine Fourier components for each of the fast and Alfv\'en potentials $\Phi$ and $\Psi$ respectively. Again it is clear that only the $m=0$, $n=-1$ mode exhibits a significant travelling fast wave, with $\Phi$ essentially becoming evanescent in the other modes as the top is approached. The travelling nature of the Alfv\'en wave is apparent in the $\Psi$ figures. Take particular note of the decaying Alfv\'en wave in $m=n=0$.

\begin{figure}
\begin{center}
\includegraphics[width=.98\hsize]{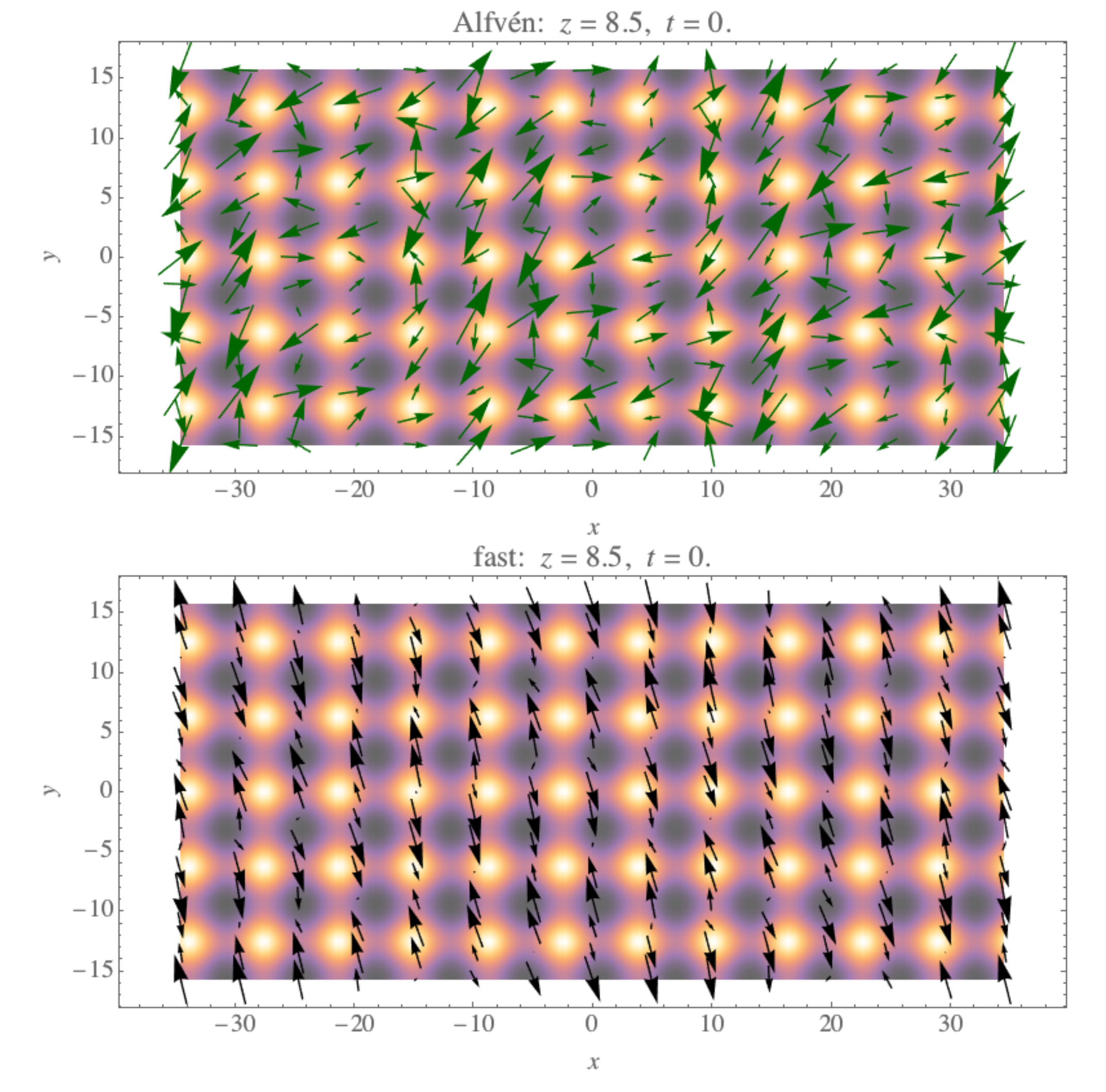}
\caption{Displacement vectors at a particular time for the case of Figure \ref{fig:kase4spec}, corresponding to the Alfv\'en component (upper panel) and the fast component (lower panel). The $y$-component of each plotted vector is indeed $\xi_y$, but the $x$ component is $\xi_\perp$ rather than $\xi_x$ for purposes of display. That is, these are the displacements seen along the line of the magnetic field, though at fixed height $z=8.5$. Arrow lengths represent the true displacement comparisons. Two animations, of the Alfv\'en and fast displacements respectively, accompany this paper.}
\label{fig:kase4xi}
\end{center}
\end{figure}

\begin{figure}
\begin{center}
\includegraphics[width=\hsize]{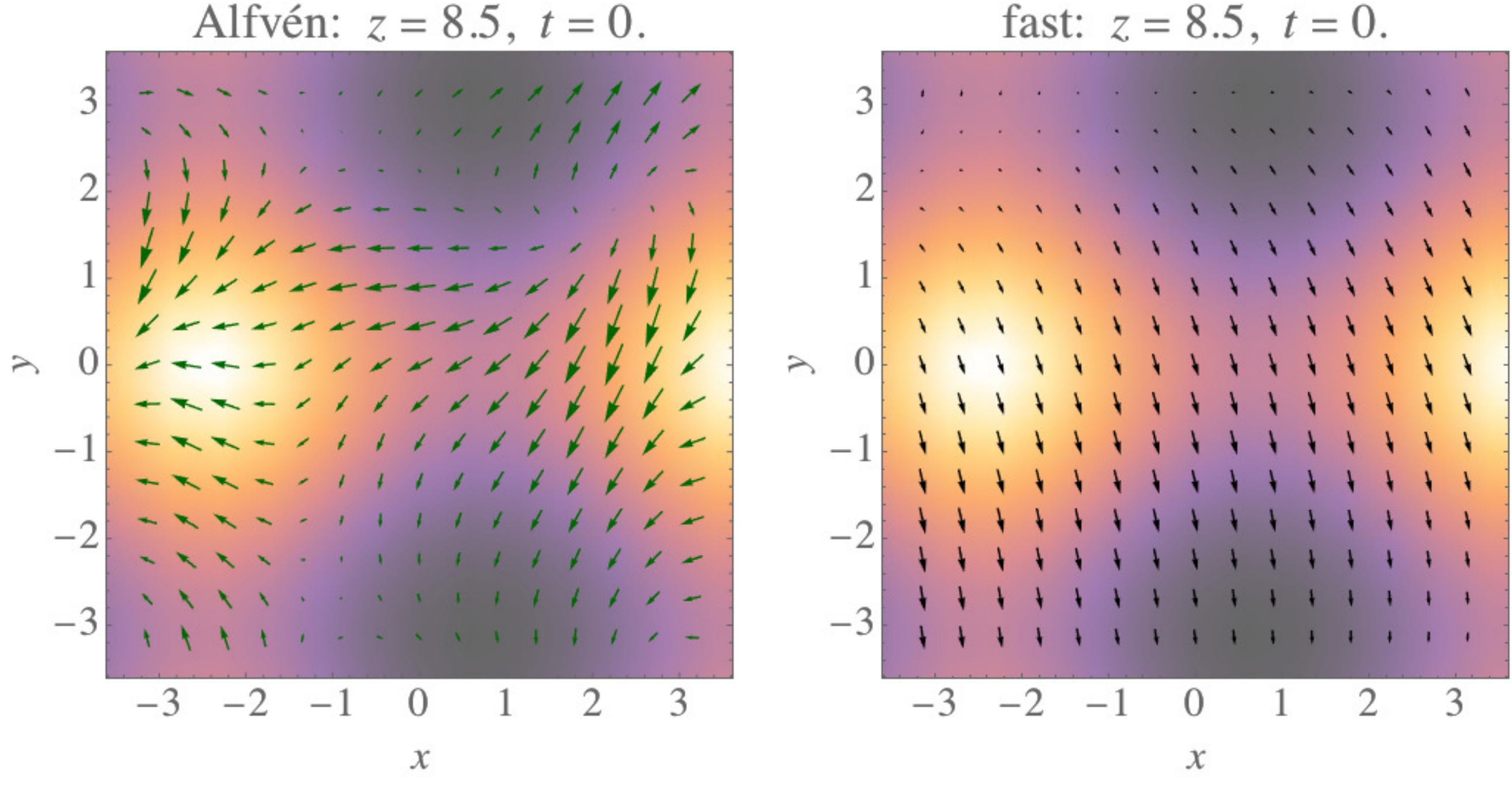}
\caption{Zoomed version of Figure \ref{fig:kase4xi}, showing Alfv\'en (left) and fast (right) displacements.}
\label{fig:kase4xizoom}
\end{center}
\end{figure}

Figure \ref{fig:kase4xi} and \ref{fig:kase4xizoom} show snapshots of the displacement vector $\bxi$ associated with the Alfv\'en and fast waves independently. As is to be expected in light of the flux comparisons, the Alfv\'en displacements are significantly larger than those associated with the fast wave. Watching a movie of these displacements confirms that the Alfv\'en displacements predominantly rotate, whereas the fast displacements are approximately linear and therefore to be identified with kink-type (transverse) waves. Figure \ref{fig:kase4xizoom} in particular confirms the earlier conclusion that the sole propagating fast wave for this case ($m=0$, $n=-1$) is space-filling, and not restricted to tubes.

\begin{figure}
\begin{center}
\includegraphics[width=\hsize]{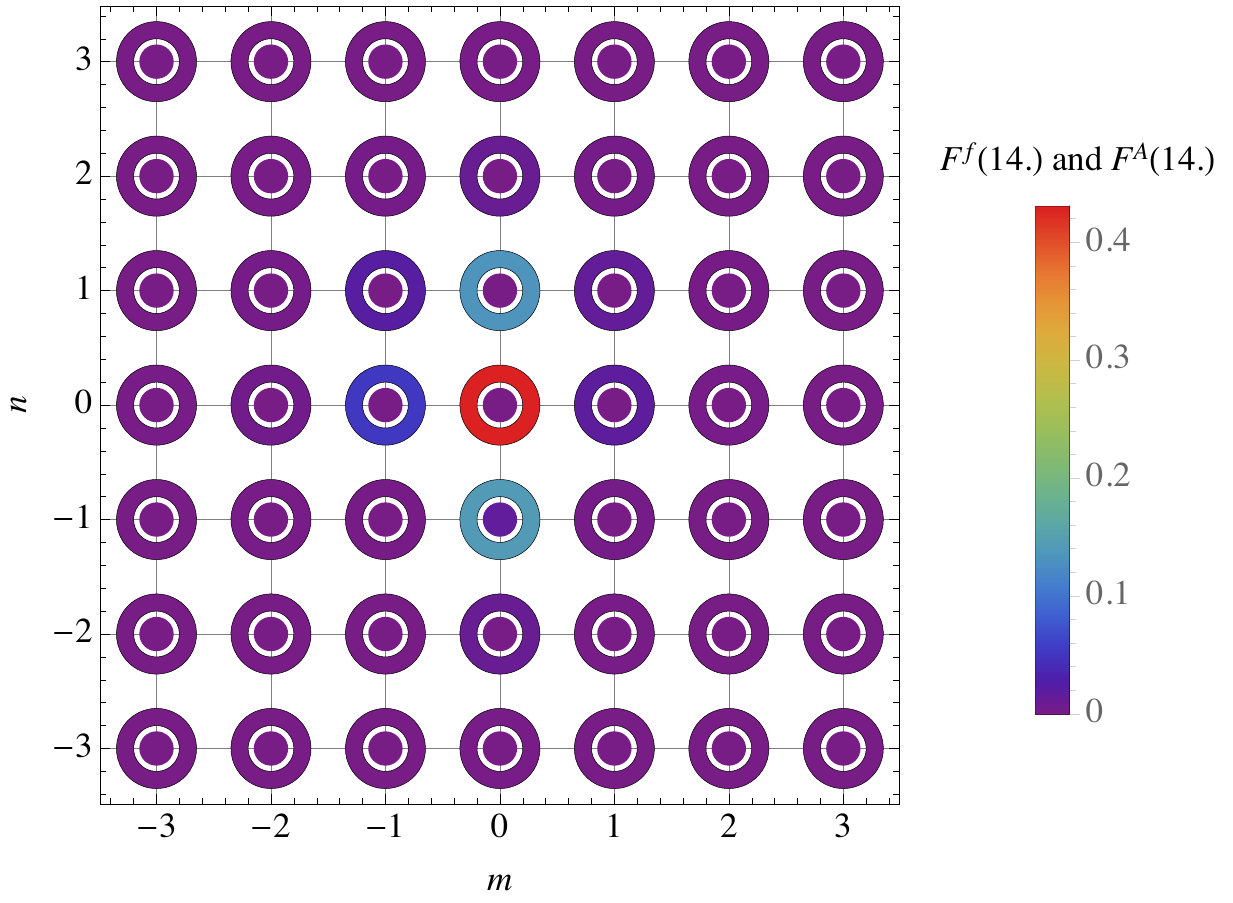}
\caption{Top Alfv\'en (outer annulus) and fast (inner disk) fluxes in the Fourier modes $m$, $n$ for the same case as in Figure \ref{fig:kase4spec}, but with less inclined magnetic field, $\theta=20^\circ$.}
\label{fig:kase5spec}
\end{center}
\end{figure}

\begin{figure}
\begin{center}
\includegraphics[width=\hsize]{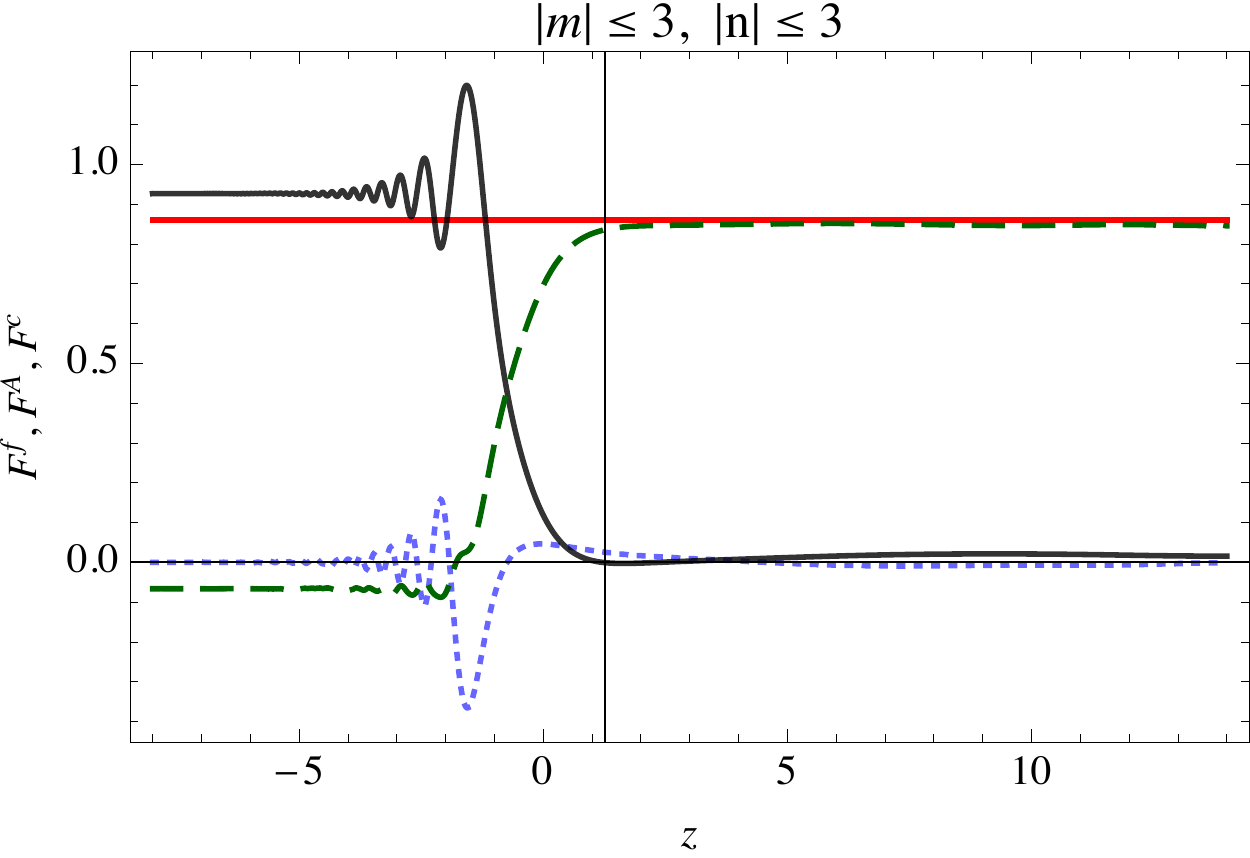}
\caption{Alfv\'en (green dashed), fast (black full), cross (dotted blue), and total (red full) fluxes summed over all Fourier modes $-3\le m,\,n\le3$ as functions of height $z$ for the case of Figure \ref{fig:kase5spec}, i.e., with field inclination $\theta=20^\circ$.}
\label{fig:kase5Ftot}
\end{center}
\end{figure}

\begin{figure}
\begin{center}
\includegraphics[width=\hsize]{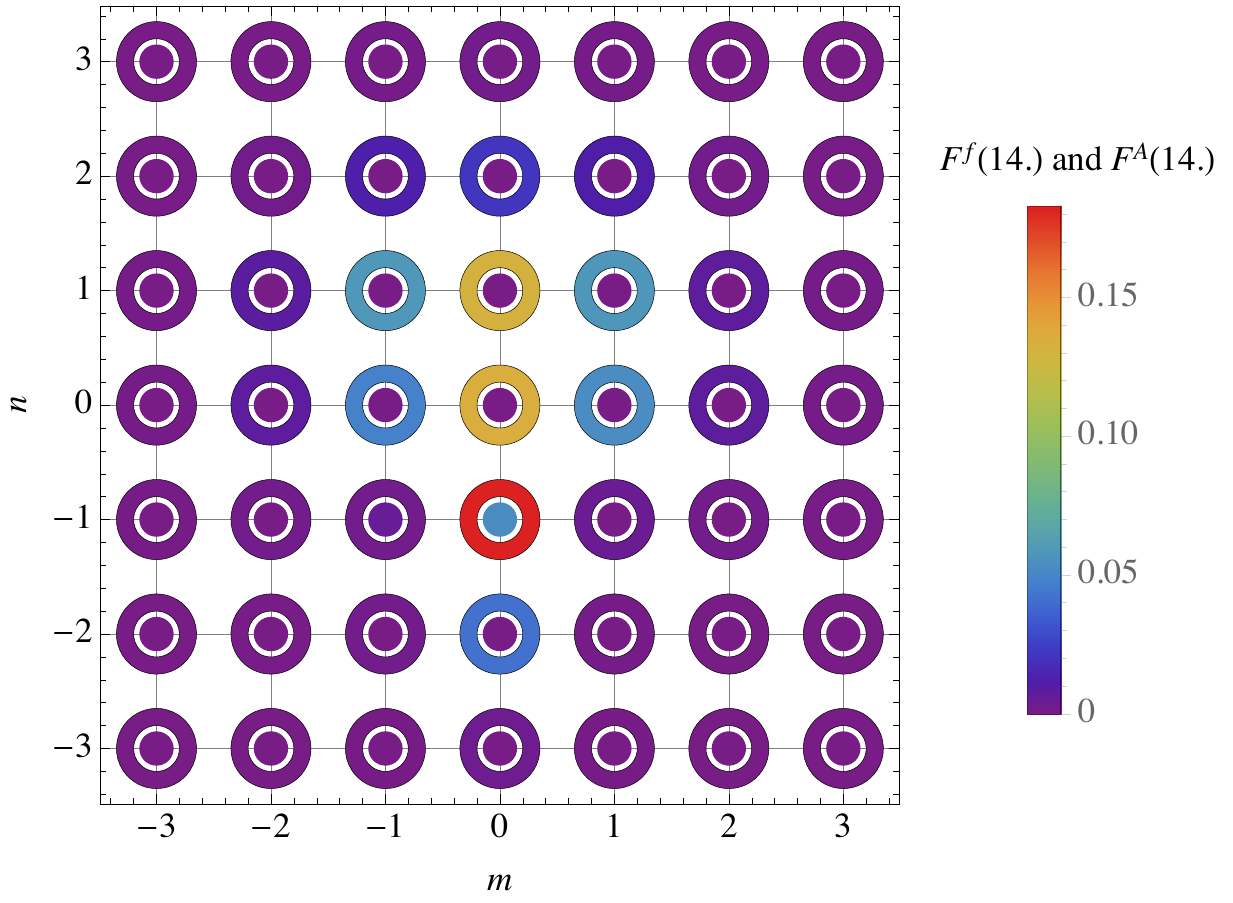}
\caption{Top Alfv\'en (outer annulus) and fast (inner disk) fluxes in the Fourier modes $m$, $n$ for the same case as in Figure \ref{fig:kase5spec} (i.e., $\theta=20^\circ$), but with $r=1/3$, $s=5/6$.}
\label{fig:kase6spec}
\end{center}
\end{figure}

Figure \ref{fig:kase5spec} shows the spectral fluxes for flux tubes inclined only $20^\circ$ from the vertical, rather than $50^\circ$ as before. Correspondingly, Figure \ref{fig:kase5Ftot} indicates that there is much-reduced (though not zero) power in the kink mode for less-inclined flux tubes. Figure \ref{fig:kase6spec} presents the spectral fluxes, again for $\theta=20^\circ$, but with a larger driving wavevector $r=1/3$, $s=5/6$, exhibiting differences in detail, but the same overall conclusions.

%%%%%
\subsubsection{$\alpha=1.15$ Swiss Cheese} \label{cheese}
Now consider a case ($1<\alpha<1+\epsilon_0$) where the region in which fast waves may propagate becomes contiguous, not restricted to discrete flux tubes. With $\alpha=1.15$ and $\epsilon_0=0.3$ for example, this occupies 69\% of the cross-sectional area. It is now Swiss cheese rather than an ensemble of separated waveguides.

\begin{figure}
\begin{center}
\includegraphics[width=\hsize]{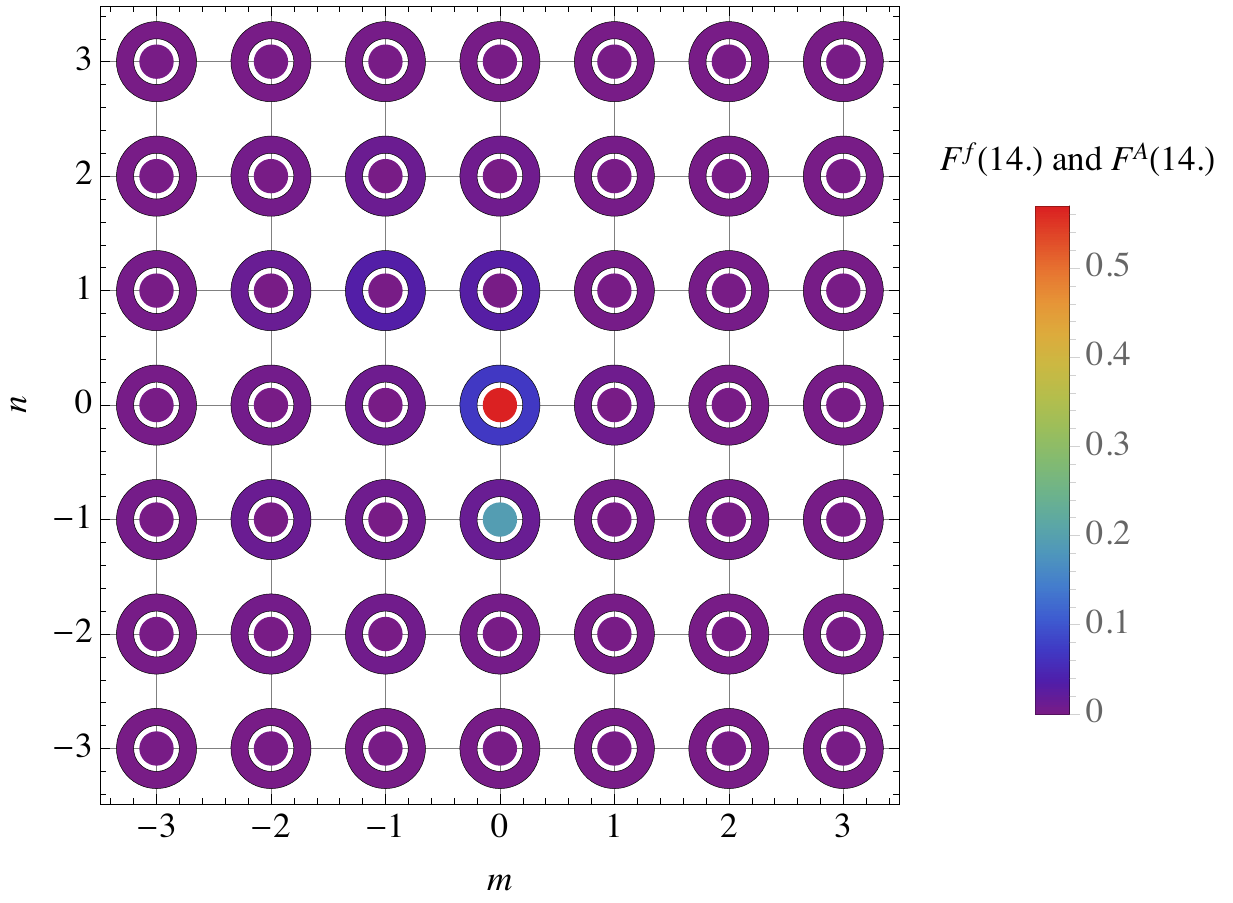}
\caption{Top Alfv\'en (outer annulus) and fast (inner disk) fluxes in the Fourier modes $m$, $n$ for the case $r=1/11$, $s=3/5$, $\theta=50^\circ$, $\epsilon_0=0.3$, $\alpha=1.15$ , $L=12$, $W=1$, $z_\text{bot}=-8$, $z_\text{top}=14$ with $-3\le m,\,n\le3$. This is the same as for the case of Figure \ref{fig:kase4spec}, except for the higher frequency ($\alpha=1.15$ rather than $\alpha=0.85$). Fast propagation is no longer restricted to discrete tubes, but occupies 69\% of the cross-sectional area with discrete non-propagating tubes cut out (Swiss cheese). }
\label{fig:kase7spec}
\end{center}
\end{figure}

Unsurprisingly, the fast wave now propagates much more freely. Figure \ref{fig:kase7spec} illustrates this for the $\alpha=1.15$ case, with all other parameters as for Figure \ref{fig:kase4spec}. Though fast mode flux now dominates, there is still considerable Alfv\'en power (27\%). In any case, the fast wave can no longer be described as a kink wave,  as there are no longer discrete wave guides. It is more a case of a propagating bulk fast wave with excisions. The modes $m=0$ with $n=0$ and $-1$ support travelling fast waves. In this case, the spectral scatter has taken otherwise propagating fast waves \emph{out of} the propagating regime, though it has left two strong channels for fast propagation.

%%%%%
\subsubsection{$\alpha=0.6$} \label{noprop}
Needless to say, when $\alpha<1-\epsilon_0$, there is essentially no fast wave power. This case has been checked numerically for the same model as for Figure \ref{fig:kase7spec} but with $\alpha=0.6$; it supports only Alfv\'en waves beyond about $z=1$. No graphs are presented here for that case, as there is nothing surprising to report.

%%%%%%%%%%%%%%%%%%%%%%%%%%%%%%%%%%%%%%%%%%%%%%%%%
\section{Conclusion}
Despite the simplicity of the model, the results presented here are instructive, and illustrate a number of features that might be expected in complex mixed vertical/cross-field structured atmospheres.

The following lessons may be drawn.
\begin{enumerate}
\item A major effect of the packed flux tube structure is to scatter in Fourier space. This can partially scatter an evanescent fast wave into travelling fast waves, primarily manifesting as kink waves, but also possibly as space-filling fast waves. Conversely, it can scatter travelling fast waves to higher wave number where they are evanescent. 
\item For the most part, it can be expected that the bulk of seismically generated fast wave flux incident from below reflects before it reaches the TR, so the process of scatter \emph{into} travelling fast/kink modes provides a mechanism for carrying fast waves upward that would not be available in an unstructured corona. The $m=0$, $n=-1$ (bottom centre) panel of Figure \ref{fig:kase4Ftot} illustrates this well, with a general build-up of fast wave flux over $2\lesssim z \lesssim 7$. The cross-flux is relatively small in this region, so an interpretation in terms of fast kink waves is justified.
\item However, the kink wave eventually starts to decay via resonant coupling to the Alfv\'en wave (see the same panel for $8\lesssim z \lesssim 12$).
\item The tubes also scatter Alfv\'en waves in Fourier space, but these are all travelling waves. This is a major distinction between fast and Alfv\'en waves in the flux tube ensemble: the former carry energy upward only for a very restricted range of wave numbers, if any, whereas the latter can do so at all $(m,n)$.
\item Alfv\'en waves themselves scatter into higher mode numbers, representing the process for mode mixing. The central panel of Figure \ref{fig:kase4Ftot} is a good example of this.
\item Alfv\'en energy will be spread more widely in Fourier space if the flux tubes are allowed to extend much higher than numerical constraints have permitted here. This will see the oscillations disappear from view in practice.
\item High transmissions through the transition region at $z=0$ are easily attained (recalling that the original incident wave carried unit flux). This is encouraging from the point of view of coronal heating and solar wind acceleration. 
\item Only incident fast waves have been considered here, in line with the supposition that these waves originate from the Sun's internal seismology. Direct injection of Alfv\'en waves at the base may be of interest, but is likely to be less realistic in the solar context because of the difficulty of generating Alfv\'en waves at the weakly ionized photosphere.
\end{enumerate}

In summary, the observable kink-like oscillations presumably responsible for the various CoMP, AIA, and SOT observations may represent only a small part of the total upward wave flux in coronal flux tube ensembles. It is notable that net upward flux in all cases explored is a significant fraction of the injected flux, so wave energies, both fast and Alfv\'en, may in combination provide ample energy to supply the corona.

%%%%%%%%%%%%%%%%%%%%%%%%%%%%%%%%%%%%%%%%%%%%%%%%%%%%%%%

%%%%%%%%%%%
%\acknowledgements

%%%%%%%%%%%%%%%%%%%%%%%%%%%%%%%%%%%%%%%%%%%%%%%%%%%%%%%%%%%%%%%%%%%%%%%%%%%
%\goodbreak

\bibliographystyle{mnras}        
\bibliography{fred}

\end{document}